\documentclass[floatfix,twocolumn,showpacs,prb,amsmath]{revtex4}
\usepackage{graphicx}
\usepackage{bm}
\usepackage{amssymb}
\usepackage{dcolumn}
\usepackage{subfigure}
\usepackage{threeparttable}
\usepackage{multirow}
\usepackage[usenames,dvipsnames]{color}
\usepackage{wasysym}
\usepackage{txfonts}
\usepackage{mathtools}

\begin{document}
\newcommand{\vn}[1]{{\bf{#1}}}
\newcommand{\vht}[1]{{\boldsymbol{#1}}}
\newcommand{\matn}[1]{{\bf{#1}}}
\newcommand{\matnht}[1]{{\boldsymbol{#1}}}
\newcommand{\bege}{\begin{equation}}
\newcommand{\ee}{\end{equation}}
\newcommand{\bal}{\begin{aligned}}
\newcommand{\defbar}{\overline}
\newcommand{\SM}{\scriptstyle}
\newcommand{\eal}{\end{aligned}}
\newcommand{\udot}{\overset{.}{u}}
\newcommand{\exponential}[1]{{\exp(#1)}}
\newcommand{\phandot}[1]{\overset{\phantom{.}}{#1}}
\newcommand{\phandag}{\phantom{\dagger}}
\newcommand{\Trace}{\text{Tr}}
\newcommand{\Bxc}{\Omega}
\title{Spin-orbit torques in Co/Pt(111) and Mn/W(001) 
magnetic bilayers from first principles}
\author{Frank Freimuth}
\email[Corresp.~author:~]{f.freimuth@fz-juelich.de}
\author{Stefan Bl\"ugel}
\author{Yuriy Mokrousov}
\affiliation{Peter Gr\"unberg Institut and Institute for 
Advanced Simulation,
Forschungszentrum J\"ulich and JARA, 52425 J\"ulich, Germany}
\date{\today}
\begin{abstract}
An applied electric current through 
a space-inversion asymmetric magnet induces
spin-orbit torques (SOTs) on the magnetic moments, which holds much
promise for future memory devices.
We discuss general Green's function expressions
suitable to compute the
linear-response SOT in disordered ferromagnets.
The SOT can be decomposed into an even and an odd component 
with respect to magnetization reversal,
where 
in the limit of vanishing disorder
the even 
SOT is given by the constant Berry curvature of the occupied states, 
while the odd part exhibits a divergence with respect
to disorder strength.
Within this formalism, we perform first principles density-functional theory 
calculations of the SOT in Co/Pt(111) and Mn/W(001) magnetic bilayers. 
We find the even and odd
torque components to be of comparable magnitude.
Moreover, the odd torque depends strongly on an 
additional capping layer, while the
even torque is less sensitive. 
We show that the even 
torque is nearly entirely mediated by spin currents in contrast to
the odd torque, which can contain an important 
contribution not due to spin transfer.
Our results are in
agreement with experiments, showing that our linear-response theory
is well-suited for the description of SOTs in complex ferromagnets.
\end{abstract}


\maketitle
\section{Introduction}
The combination of spin-orbit interaction (SOI) 
and broken inversion symmetry gives rise
to torques on the magnetization of ferromagnets if an electric 
current is applied~\cite{manchon_zhang_2009,torque_macdonald}.
These so-called spin-orbit torques (SOTs) arise from 
the exchange of angular momentum
between the crystal lattice and the magnetization 
and enable control of the magnetic state
of a single ferromagnetic layer, while conventional  
spin-transfer torque (STT)~\cite{stiles_miltat_stt} devices 
exploit the exchange of spin angular momentum
between two ferromagnetic layers with different 
magnetization directions.

SOTs can be observed experimentally both in periodic crystals --
if the structure lacks inversion symmetry like 
in bulk (Ga,Mn)As with 
zinc-blende crystalline 
structure~\cite{chernyshov_2009,sot_kurebayashi} -- and in trilayers
with structural inversion asymmetry, e.g.\ in 
AlO$_{x}$/Co/Pt~\cite{CoPtAlO_spin_torque_rashba_Gambardella,
CoPtAlO_perpendicular_switching_Gambardella,
current_induced_switching_using_spin_torque_from_spin_hall_buhrman,tilting_spin_orientation_pi} and
MgO/CoFeB/Ta~\cite{spin_torque_switching_TaCoFeB_Buhrman,
layer_thickness_dependence_current_induced_effective_field_TaCoFeB_Hayashi,
current_induced_effective_field_TaCoFeB_Suzuki},
where a ferromagnetic layer is asymmetrically sandwiched between 
a heavy-metal layer and an oxide layer and 
the applied current is parallel to the
interfaces.
The observation of magnetization
switching by SOTs in
systems with strong perpendicular anisotropy
suggests promising new ways to realize magnetic memory 
devices~\cite{CoPtAlO_perpendicular_switching_Gambardella,
current_induced_switching_using_spin_torque_from_spin_hall_buhrman}.  

Two qualitatively different SOTs are found in experiments 
on trilayers, 
one is an
even function of magnetization direction $\hat{\vn{M}}$, the other one 
an odd function. 
To lowest order in $\hat{\vn{M}}$, they are given by
$\vn{T}^{\rm even}=
{\rm t}^{\rm even}
\hat{\vn{M}}\times[(\hat{\vn{e}}_{z}\times\vn{E})\times\hat{\vn{M}}]$
and $\vn{T}^{\rm odd}=
{\rm t}^{\rm odd}
(\hat{\vn{e}}_{z}\times\vn{E})\times\hat{\vn{M}}$,
where $\vn{E}$ denotes the applied in-plane electric field  
and $\hat{\vn{e}}_{z}$ is a unit vector in the out-of-plane 
direction, i.e.,
perpendicular to the interfaces.
Additional higher order terms describing the anisotropy of SOTs
have been shown to be important in 
AlO$_{x}$/Co/Pt~\cite{symmetry_spin_orbit_torques}.
Inverse spin-pumping~\cite{spin_pumping_tserkovnyak} driven by the 
spin current due to the spin Hall effect (SHE) in the heavy-metal 
layer is expected to provide an
important contribution to $\vn{T}^{\rm even}$.
Accordingly, it has been proposed to use materials 
with large spin Hall angles
to achieve strong 
SOTs~\cite{perfect_alloys_spin_hall,spin_torque_switching_TaCoFeB_Buhrman}.
In trilayers, $\vn{T}^{\rm odd}$ 
behaves like a field-like 
torque due to an effective magnetic field 
$\propto\hat{\vn{e}}_{z}\times\vn{E}$. 
Since $\hat{\vn{e}}_{z}\times\vn{E}$ is also
the direction of the Rashba spin-orbit field for charge 
carriers moving along $\vn{E}$ in the structure
inversion asymmetric geometry,
one possible origin is the current-induced non-equilibrium 
spin accumulation along the spin-orbit field~\cite{edelstein},
which results -- via the exchange interaction -- in this
effective magnetic field~\cite{manchon_zhang_2009,CoPt_Haney_Stiles}. 
Indeed, the Rashba effect has been found to be 
strong at magnetic heavy metal surfaces and 
interfaces~\cite{rashba_magnetic_metal_surface,rashba_interaction_park}. 
Additionally, it is expected that part of $\vn{T}^{\rm odd}$ arises 
from SHE and that
part of $\vn{T}^{\rm even}$ arises from the 
Rashba effect~\cite{quantum_kinetic_rashba_macdonald,quantum_kinetic_rashba_manchon,
semi_classical_modelling_stiles}.

Understanding the roles played by the various mechanisms proposed 
to explain SOTs is crucial for optimizing
and finetuning their properties for future SOT-based devices.
For this purpose we investigate in this work 
SOTs in Mn/W(001) and Co/Pt(111) 
bilayers as well as 
in O/Co/Pt(111) and Al/Co/Pt(111) trilayers 
using Kubo linear-response calculations based on the 
first-principles electronic structure obtained from
density functional theory (DFT). 
Within a
model description of disorder we study the dependence of SOTs
on disorder strength.
Additionally, we explore the dependence of 
the SOTs on the heavy metal layer thickness.
Furthermore, we determine to what 
extent spin currents mediate the torques
by computing spin fluxes and decomposing the 
torque into contributions of individual atoms. By comparing
SOTs in Mn/W(001) and Mn/W(001)/Mn we investigate
the influence of an additional Mn substrate layer on the SOT.
Finally, we compare the spin currents that contribute to the 
magnetic anisotropy torque
to the spin currents that contribute to the SOT.

This article is structured as follows. We start with a discussion of the
linear-response formalism used for the calculation of the SOT 
and specify the computational details in
section II. Section III presents the results.
Atom-resolved torques and spin fluxes are introduced and investigated
in subsection III.B.
We conclude by a summary in section IV.
Additional derivations and aspects related to the Kubo linear-response
formalism for the SOT are given in the appendices.

\section{Computational method}
\subsection{Kubo linear-response formalism for the torkance tensor}
Within the local spin density approximation to DFT the Hamiltonian $H$
can be decomposed as~\cite{noco_flapw}
\bege\label{eq_hamiltonian}
H=H_{0}+\mu_{\rm B}\vht{\sigma}\cdot\vn{\Bxc}^{\rm xc},
\ee
where $H_{0}$ contains kinetic energy, scalar potential and SOI.
$\mu_{\rm B}$ is the Bohr magneton, 
$\vht{\sigma}=(\sigma_x,\sigma_y,\sigma_z)^{\rm T}$ is the 
vector of Pauli spin matrices,
and $\vn{\Bxc}^{\rm xc}$ is the exchange field.
We consider only ferromagnetic systems, where the exchange
field $\vn{\Bxc}^{\rm xc}(\vn{r})=\Bxc^{\rm xc}(\vn{r})\hat{\vn{M}}$
is characterized by a
position-independent direction $\hat{\vn{M}}$ and
a position-dependent amplitude $\Bxc^{\rm xc}(\vn{r})$.
The relation to the Kohn-Sham effective
potentials $V^{\rm eff}_{\rm majority}(\vn{r})$
and $V^{\rm eff}_{\rm minority}(\vn{r})$ of majority and
minority electrons is given by
$\Bxc^{\rm xc}(\vn{r})=\frac{1}{2\mu_{\rm B}}\left[
V^{\rm eff}_{\rm minority}(\vn{r})-V^{\rm eff}_{\rm majority}(\vn{r})
\right]$.
In response to an applied electric field a
magnetization $\delta \vn{M}(\vn{r})$ is
induced at position $\vn{r}$. As a consequence,
the exchange field $\vn{\Bxc}^{\rm xc}(\vn{r})$ is modified by
$\delta\vn{\Bxc}^{\rm xc}(\vn{r})=
\Bxc^{\rm xc}(\vn{r})
\delta \vn{M}(\vn{r}) 
/M(\vn{r})$.
The resulting torque $\vn{T}$ on the magnetization 
within one unit cell is given
by~\cite{stt-cocuco_haney&waldron&duine&nunez&guo&macdonald,
CurrentinducedtorquesinmagneticmetalsBeyondspintransfer_Haney_2008}
\bege
\vn{T}=
\int d^{3}r \vn{M}(\vn{r})\times \delta\vn{\Bxc}^{\rm xc}(\vn{r})=
\int d^{3}r \vn{\Bxc}^{\rm xc}(\vn{r})\times \delta \vn{M}(\vn{r}),
\ee
where the integration is over the unit cell volume.
Thus, the torque on the magnetization arises from the component of
$\delta \vn{M}(\vn{r})$ that is perpendicular to $\vn{\Bxc}^{\rm xc}(\vn{r})$.
Within linear-response theory 
the torque $\vn{T}$ arising due
to an applied electric field $\vn{E}$
can be written as $\vn{T}=\vn{t}\vn{E}$, 
which defines the torkance tensor $\vn{t}$.
From the
Kubo formalism we derive the expression 
${\rm t}^{\phantom{II}}_{ij}={\rm t}^{\rm I(a)}_{ij}+
{\rm t}^{\rm I(b)}_{ij}+
{\rm t}^{\rm II}_{ij}
$,
where
(see Appendix~\ref{app_sot_formula})
\begin{gather}\label{eq_kubo_linear_response}
\begin{aligned}
{\rm t}^{\rm I(a)\phantom{I}}_{ij}\!\!\!\!&=\phantom{-}\frac{e}{h}
\,{\rm Tr}
\left\langle
\mathcal{T}_{i}
G^{\rm R}(\mathcal{E}_{\rm F})
v_{j}
G^{\rm A}(\mathcal{E}_{\rm F})
\right\rangle
\\
{\rm t}^{\rm I(b)\phantom{I}}_{ij}\!\!\!\!&=-\frac{e}{h}
\,{\rm Re}
\,{\rm Tr}
\left\langle
\mathcal{T}_{i}
G^{\rm R}(\mathcal{E}_{\rm F})
v_{j}
G^{\rm R}(\mathcal{E}_{\rm F})
\right\rangle
\\
{\rm t}^{\rm II\phantom{(a)}}_{ij}\!\!\!\!&=
\phantom{-}\frac{e}{h}\int_{-\infty}^{\mathcal{E}_{\rm F}}
d\mathcal{E}
\,{\rm Re}
\,{\rm Tr}
\left\langle
\mathcal{T}_{i}G^{\rm R}(\mathcal{E})v_{j}
\frac{dG^{\rm R}(\mathcal{E})}{d\mathcal{E}}\right.\\
 &\quad\quad\quad\quad\quad\quad\quad\quad\,-\left.
\mathcal{T}_{i}\frac{dG^{\rm R}(\mathcal{E})}{d\mathcal{E}}v_{j}G^{\rm R}(\mathcal{E})
\right\rangle.
\end{aligned}
\end{gather}
Here, $G^{\rm R}(\mathcal{E})$ 
is the retarded Green function,
$G^{\rm A}(\mathcal{E})$ is the  
advanced one,
$\mathcal{E}_{\rm F}$ is the Fermi energy, 
$e>0$ is the elementary positive charge and
$v_{i}$ is the $i$-th cartesian
component of the velocity operator. 
The torque operator at position $\vn{r}$ is given by
$\vht{\mathcal{T}}(\vn{r})
=-\mu_{\rm B}\vht{\sigma}\times\vn{\Bxc}^{\rm xc}(\vn{r})$, 
and
$\mathcal{T}_{i}$ is its $i$-th cartesian component.

In order to compare theory with experiment, 
we decompose the computed torkance into
its even and odd 
parts: $\vn{t}(\hat{\vn{M}})
=\vn{t}^{\rm even}(\hat{\vn{M}})
+\vn{t}^{\rm odd}(\hat{\vn{M}})$,
where $\vn{t}^{\rm even}(\hat{\vn{M}})
=[\vn{t}(\hat{\vn{M}})
+\vn{t}(-\hat{\vn{M}})]/2$ and 
$\vn{t}^{\rm odd}(\hat{\vn{M}})
=[\vn{t}(\hat{\vn{M}})-\vn{t}(-\hat{\vn{M}})]/2$.
The same decomposition into even and odd parts 
is widely used in the
case of the conductivity 
tensor \mbox{$\sigma^{\phantom{\rm e}}_{ij}(\hat{\vn{M}})$} 
(see also Eq.~\eqref{eq_kubo_linear_response_conductivity}):
$\sigma^{\phantom{\rm e}}_{ij}(\hat{\vn{M}})
=\sigma_{ij}^{\rm even}(\hat{\vn{M}})
+\sigma_{ij}^{\rm odd}(\hat{\vn{M}})$,
where $\sigma_{ij}^{\rm even}(\hat{\vn{M}})
=[\sigma_{ij}(\hat{\vn{M}})
+\sigma_{ij}(-\hat{\vn{M}})]/2$ and 
$\sigma_{ij}^{\rm odd}(\hat{\vn{M}})
=[\sigma_{ij}(\hat{\vn{M}})-\sigma_{ij}(-\hat{\vn{M}})]/2$.
Due to the Onsager 
relation $\sigma_{ij}(\hat{\vn{M}})=\sigma_{ji}(-\hat{\vn{M}})$
the even part of the conductivity tensor is symmetric, i.e.,
$\sigma_{ij}^{\rm even}(\hat{\vn{M}})=\sigma_{ji}^{\rm even}(\hat{\vn{M}})$,
while the odd part of the conductivity tensor is antisymmetric, i.e.,
$\sigma_{ij}^{\rm odd}(\hat{\vn{M}})=-\sigma_{ji}^{\rm odd}(\hat{\vn{M}})$.~\cite{birss} 
However, in the case of the SOTs, the Onsager reciprocity dictates that
a time-dependent magnetization $\hat{\vn{M}}(t)$
induces a current density $\vn{j}(t)=[\vn{t}(-\hat{\vn{M}}(t))]^{\rm T}
[\hat{\vn{M}}(t)\times\frac{d \hat{\vn{M}}(t)}{dt}]/V$,
where $V$ is the unit cell volume.~\cite{invsot}
Thus, while the Onsager reciprocity relates different matrix elements of
the conductivity tensor, it does not relate different matrix elements of
the torkance tensor, but instead relates the SOT to its inverse. Consequently,
the even torkance is in general neither symmetric nor antisymmetric and
likewise the odd torkance is in general neither symmetric nor antisymmetric.

We approximate the effect of disorder by 
a constant band broadening $\Gamma$, i.e.,
by setting $G^{\rm R}(\mathcal{E})=\hbar[\mathcal{E}-H+i\Gamma]^{-1}$,
where $H$ is the Hamiltonian Eq. ~(1).~\cite{prb_she_tanaka}
In the case of anomalous Hall effect (AHE) 
and SHE this constant
$\Gamma$ approximation does not capture 
the so-called side-jump and
skew-scattering~\cite{ahe_ebert,ahe_turek,she_mertig,sj_sinova,sj_juergen}.
The computational assessment of formally
analogous extrinsic contributions to the 
torkance is not considered here and 
left for future work.
Within the constant $\Gamma$ approximation the even torkance is 
given by (see Appendix~\ref{app_sot_formula})
\bege
\begin{aligned}
\label{eq_even_torque_constant_gamma}
{\rm t}^{\rm even}_{ij}=&
\frac{e\hbar}{2\pi\mathcal{N}}
\sum_{\vn{k}n\ne m}
{\rm Im}
\left[ 
\langle 
\psi^{\phantom{R}}_{\vn{k}n}  
|
\mathcal{T}_{i}
| 
\psi^{\phantom{R}}_{\vn{k}m}  
\rangle
\langle 
\psi^{\phantom{R}}_{\vn{k}m}  
|
v_{j}
| 
\psi^{\phantom{R}}_{\vn{k}n}  
\rangle
\right]\Biggl\{\\
&\frac{\Gamma
(\mathcal{E}^{\phantom{R}}_{\vn{k}m}
-
\mathcal{E}^{\phantom{R}}_{\vn{k}n})
}{
\left[(\mathcal{E}^{\phantom{R}}_{\rm F}
-
\mathcal{E}^{\phantom{R}}_{\vn{k}n})^2+\Gamma^2\right]
\left[(\mathcal{E}^{\phantom{R}}_{\rm F}
-
\mathcal{E}^{\phantom{R}}_{\vn{k}m})^2+\Gamma^2\right]
}+\\
+&
\frac{
2\Gamma
}
{
\left[
\mathcal{E}^{\phantom{R}}_{\vn{k}n}
-
\mathcal{E}^{\phantom{R}}_{\vn{k}m}
\right]
\left[(\mathcal{E}^{\phantom{R}}_{\rm F}
-
\mathcal{E}^{\phantom{R}}_{\vn{k}m})^2+\Gamma^2\right]
}+\\
+&
\frac{
2
}
{
\left[
\mathcal{E}^{\phantom{R}}_{\vn{k}n}
-
\mathcal{E}^{\phantom{R}}_{\vn{k}m}
\right]^2
}
{\rm Im}\,{\rm ln}
\frac{
\mathcal{E}^{\phantom{R}}_{\vn{k}m}
-
\mathcal{E}^{\phantom{R}}_{\rm F}-i\Gamma
}
{
\mathcal{E}^{\phantom{R}}_{\vn{k}n}
-
\mathcal{E}^{\phantom{R}}_{\rm F}-i\Gamma
}\Biggl\}
\end{aligned}
\ee
and the odd torkance is given by
\bege
\label{eq_odd_torque_constant_gamma}
{\rm t}^{\rm odd}_{ij}=
\frac{e\hbar}{\pi\mathcal{N}}
\sum_{\vn{k}nm}
\frac{\Gamma^2
{\rm Re}
\left[ 
\langle 
\psi^{\phantom{R}}_{\vn{k}n}  
|
\mathcal{T}_{i}
| 
\psi^{\phantom{R}}_{\vn{k}m}  
\rangle
\langle 
\psi^{\phantom{R}}_{\vn{k}m}  
|
v_{j}
| 
\psi^{\phantom{R}}_{\vn{k}n}  
\rangle\right]
}{
\left[(\mathcal{E}^{\phantom{R}}_{\rm F}-\mathcal{E}^{\phantom{R}}_{\vn{k}n})^2+\Gamma^2\right]
\left[(\mathcal{E}^{\phantom{R}}_{\rm F}-\mathcal{E}^{\phantom{R}}_{\vn{k}m})^2+\Gamma^2\right]
},
\ee
where $\mathcal{N}$ is the number of $\vn{k}$-points used 
to sample the Brillouin zone and
$\psi^{\phantom{R}}_{\vn{k}n}$ and $\mathcal{E}^{\phantom{R}}_{\vn{k}n}$ 
denote the Bloch function 
for band $n$ at $\vn{k}$ and the 
corresponding band energy, respectively.

It is instructive to consider the 
limit of $\Gamma\rightarrow 0$, where we 
obtain for the even torkance (see Appendix~\ref{app_sot_formula})
\bege
\label{eq_even_torque_clean_limit}
{\rm t}^{\rm even}_{ij}\overset{\Gamma\rightarrow 0}{=}
\frac{2e}{\mathcal{N}}
\hat{\vn{e}}_{i} \cdot
\sum_{\vn{k}}\sum_{n}^{\rm occ}
\left[
\hat{\vn{M}}\times 
{\rm Im}
\left\langle 
\frac{\partial u^{\phantom{R}}_{\vn{k}n}}{\partial\hat{\vn{M}}}
\left| 
\frac{\partial u^{\phantom{R}}_{\vn{k}n}}{\partial k_{j}}\right.   
\right\rangle \right]
\ee
and for the odd torkance
\bege\label{eq_odd_torque_clean_limit}
{\rm t}^{\rm odd}_{ij}\overset{\Gamma\rightarrow 0}{=}
\frac{e\hbar}{2\Gamma\mathcal{N}}
\sum_{\vn{k}n}
\langle
\psi^{\phantom{R}}_{\vn{k}n}
|
\mathcal{T}_{i}
|
\psi^{\phantom{R}}_{\vn{k}n}
\rangle
\langle
\psi^{\phantom{R}}_{\vn{k}n}
|
v_{j}
|
\psi^{\phantom{R}}_{\vn{k}n}
\rangle
\delta(\mathcal{E}^{\phantom{R}}_{\rm F}
-
\mathcal{E}^{\phantom{R}}_{\vn{k}n}).
\ee
In Eq.~\eqref{eq_even_torque_clean_limit} $\hat{\vn{e}}_{i}$ denote 
the unit vectors
in $x$, $y$ and $z$ direction, where $\hat{\vn{e}}_{x}$ and
$\hat{\vn{e}}_{y}$ are in the plane of the trilayers,
the summation over 
band index $n$ is restricted to the occupied (occ) states
and $u^{\phantom{R}}_{\vn{k}n}(\vn{r})
=e^{-i\vn{k}\cdot\vn{r}}
\psi^{\phantom{R}}_{\vn{k}n}(\vn{r})$ is the
lattice periodic part of the Bloch 
function $\psi^{\phantom{R}}_{\vn{k}n}(\vn{r})$.
$\vn{t}^{\rm even}$ is independent 
of $\Gamma$ in the limit $\Gamma\rightarrow 0$ and describes 
the intrinsic contribution to 
the torkance. Like the 
intrinsic AHE~\cite{rmp_ahe} it has the form of a Berry curvature.
This Berry curvature contribution to the SOT was recently observed
in (Ga,Mn)As.~\cite{sot_kurebayashi}
In contrast to $\vn{t}^{\rm even}$, $\vn{t}^{\rm odd}$ diverges 
like $\Gamma^{-1}=2\tau/\hbar$ in the
limit $\Gamma\rightarrow 0$, i.e., proportional to the 
relaxation time $\tau$. 
A recent first principles study~\cite{CoPt_Haney_Stiles} 
addressed the $\Gamma\rightarrow 0$ limit
of $\vn{t}^{\rm odd}$ in the case of Co/Pt bilayers.

The Eqs.~\eqref{eq_even_torque_constant_gamma} 
and \eqref{eq_odd_torque_constant_gamma} are used in 
section~\ref{sec_results} to evaluate the SOT based on
$\mathcal{T}_{\vn{k}i nm}=\langle 
\psi^{\phantom{R}}_{\vn{k}n}  
|
\mathcal{T}_{i}
| 
\psi^{\phantom{R}}_{\vn{k}m}  
\rangle$,
$v_{\vn{k}i nm}=\langle 
\psi^{\phantom{R}}_{\vn{k}n}  
|
v_{i}
| 
\psi^{\phantom{R}}_{\vn{k}m}  
\rangle$,
$\mathcal{E}^{\phantom{R}}_{\vn{k}m}$
and
$\mathcal{E}^{\phantom{R}}_{\rm F}$
obtained from first principles electronic structure calculations.
In order to converge the $\vn{k}$ summations in these expressions
numerically efficiently, we made use of the Wannier interpolation
technique~\cite{wannierinterpolation}. Therefore, 
we first set up matrix elements 
of the necessary operators in the 
basis of maximally localized Wannier 
functions (MLWFs)~\cite{rmp_wannier90}, i.e., we 
compute 
$\langle
W_{n\vn{0}}
|
H
|
W_{m\vn{R}}
\rangle$
and
$\langle
W_{n\vn{0}}
|
\mathcal{T}_{i}
|
W_{m\vn{R}}
\rangle$,
where $|W_{m\vn{R}}\rangle$ are the MLWFs.
In order to obtain $\mathcal{T}_{\vn{k}i nm}$,
$v_{\vn{k}i nm}$ and $\mathcal{E}_{\vn{k}m}$
at a given $\vn{k}$-point we compute the Fourier transformations
\bege
\begin{aligned}
H^{\rm (W)}_{\vn{k}nm}&=\sum_{\vn{R}}
e^{i\vn{k}\cdot\vn{R}}
\langle
W_{n\vn{0}}
|
H
|
W_{m\vn{R}}
\rangle\\
v^{\rm (W)}_{\vn{k}i nm}&=\frac{1}{\hbar}\sum_{\vn{R}}
e^{i\vn{k}\cdot\vn{R}}i{\rm R}_{i}
\langle
W_{n\vn{0}}
|
H
|
W_{m\vn{R}}
\rangle\\
\mathcal{T}^{\rm (W)}_{\vn{k}i nm}&=\sum_{\vn{R}}
e^{i\vn{k}\cdot\vn{R}}
\langle
W_{n\vn{0}}
|
\mathcal{T}_{i}
|
W_{m\vn{R}}
\rangle\\
\end{aligned}
\ee
and transform them into the eigenstate representation according to
\bege
\begin{aligned}
v^{\phantom{W}}_{\vn{k}i nm}&=
\sum_{n'm'}
U_{\vn{k}n'n}^*
v^{\rm(W)}_{\vn{k}i n'm'}
U_{\vn{k}m'm}^{\phantom{*}}
\\
\mathcal{T}^{\phantom{W}}_{\vn{k}i nm}&=
\sum_{n'm'}
U_{\vn{k}n'n}^*
\mathcal{T}^{\rm(W)}_{\vn{k}i n'm'}
U_{\vn{k}m'm}^{\phantom{*}},
\end{aligned}
\ee
where the columns of the matrix $U_{\vn{k}}$ are the eigenvectors of
$H_{\vn{k}}$:
\bege
\sum_{m'} H_{\vn{k}nm'}U_{\vn{k}m'm}
=
\mathcal{E}_{\vn{k}m}U_{\vn{k}nm}.
\ee

\subsection{Computational details}
We performed DFT calculations of the electronic
structure of Mn/W(001), where we considered a 
monolayer of Mn on 9 atomic layers of W 
(denoted in the following as Mn(1)/W(9))
and additionally a monolayer of Mn on 15 atomic 
layers of W (Mn(1)/W(15)).
In order to investigate the effect of a 
second ferromagnetic 
Mn layer on the torque 
we also studied a Mn(1)/W(9)/Mn(1) trilayer, 
where 9 atomic 
layers of W(001) are symmetrically 
sandwiched between Mn monolayers
on both sides.
Structural parameters of the Mn/W(001) interface 
have been chosen 
as determined in Ref.~\cite{MnW001_structure}.  
The asymmetric 
slabs were calculated with the film mode of 
the full-potential linearized augmented-plane-wave
program {\tt FLEUR}~\cite{fleurcode}, 
which explicitly treats the
vacuum region~\cite{Krakauer_film_mode}. 
The plane wave cutoff was set to 4.1~$a_{0}^{-1}$, the
muffin-tin (MT) radius of 
2.42 $a_{0}$ ($a_{0}$ is Bohr's radius) 
was used for both Mn and W, and the generalized gradient
approximation~\cite{PerdewBurkeEnzerhof} was employed.
We used a 24$\times$24 
Monkhorst-Pack~\cite{monkhorst_pack_mesh} $\vn{k}$-mesh 
to sample the Brillouin zone
in the selfconsistent DFT calculations 
and treated spin-orbit 
interaction within second 
variation~\cite{second_variation_soi}.

\begin{figure}
\includegraphics[height=8cm,angle=-90]{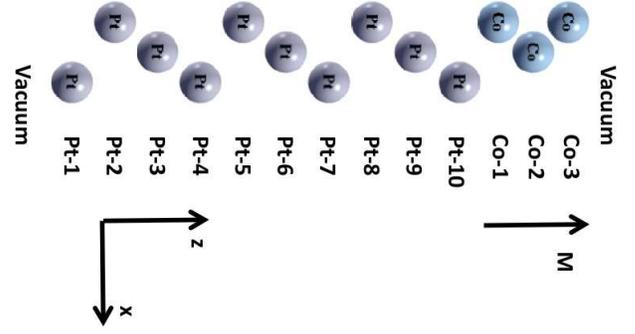}
\caption{\label{fig_unit_cell}
Illustration of the unit cell used in the thin film calculations
of the Co/Pt(111) systems. 
Magnetization $\vn{M}$
is in $z$ direction.
}
\end{figure}

In the case of Co/Pt(111) bilayers we considered 
3 layers of Co on 
7 (Co(3)/Pt(7)), 10 (Co(3)/Pt(10)), 13 (Co(3)/Pt(13)), 
15 (Co(3)/Pt(15)) and 20 (Co(3)/Pt(20)) 
atomic layers of Pt(111),
corresponding to 1.6nm, 2.3nm, 3.0nm, 3.4nm 
and 4.5nm of Pt(111), 
respectively.
We chose the (111) orientation for the fcc Pt layer
since sputter deposited Pt typically shows a strong (111)
texture along the growth 
direction.~\cite{PMA_PdCo_PtCo_thinfilms_Carcia}
In order to estimate roughly to what 
extent an oxide layer on 
the Co -- like the AlO$_{x}$ layer typically
present in experiments -- might influence the SOTs, 
we also considered
thin films composed of 10 atomic layers
of Pt(111), 3 atomic layers of Co and one additional 
atomic layer of O or Al (denoted O(1)/Co(3)/Pt(10) 
and Al(1)/Co(3)/Pt(10), 
respectively). 
The in-plane lattice constant 
of the hexagonal unit cell was set to the experimental
value of bulk Pt(111) of $a/\sqrt{2}=$~5.24~$a_0$, where
$a=$~7.41~$a_0$ is the lattice constant of the 
corresponding cubic fcc unit cell of bulk Pt. 
The corresponding 
distance between Pt atomic layers along
the (111) direction is $a/\sqrt{3}=$~4.283~$a_0$. For the Co layer 
we assumed hcp stacking and that
the first two atomic layers of Co follow the fcc pattern of
Pt(111)~\cite{CoPt_Freeman,ultrathin_CoO_on_Pt111}, i.e., 
the stacking sequence in the Pt layer
is ABC and the Co layer is stacked like ABAB onto 
the Pt-layer with termination
ABC.  
The unit cell used in the calculations is 
illustrated for the case of Co(3)/Pt(10) in Fig.~\ref{fig_unit_cell}.
The O and Al atoms are deposited at the same in-plane position as Co-2.
The plane wave cutoff was set to 3.7$a_{0}^{-1}$ and the 
following MT radii were used: 
2.5~$a_{0}$ for  Pt, 
1.8~$a_{0}$ for Co, 
1.5~$a_{0}$ for Al 
and 
1.1~$a_{0}$ for O.
Interfaces were
relaxed in the out-of-plane direction.
The resulting distance between the Pt-10 and Co-1 layers 
is 3.89~$a_0$,
while the distance between adjacent Co layers is 3.50~$a_0$.
The distances between the Co-3 and the optional Al and O capping layers
are 3.82~$a_0$ and 1.84~$a_0$, respectively.

In order to 
evaluate the Eqs.~\eqref{eq_even_torque_constant_gamma} 
and \eqref{eq_odd_torque_constant_gamma}
computationally 
efficiently we made
use of the Wannier interpolation 
technique. 
We constructed 18 MLWFs 
per transition metal atom, 
and additionally 8 MLWFs per O and Al atom using 
an 8$\times$8 $\vn{k}$ mesh.~\cite{WannierPaper,wannier90}
The subspace of the MLWFs was disentangled~\cite{disentanglement} 
from a number of bands
of 1.4 times the number of desired MLWFs. 
A 1024$\times$1024 Monkhorst-Pack~\cite{monkhorst_pack_mesh} $\vn{k}$ mesh 
was used in the Wannier interpolation 
of Eqs.~\eqref{eq_even_torque_constant_gamma} 
and \eqref{eq_odd_torque_constant_gamma}.

\begin{figure}
\subfigure[\label{fig_effective_fields_even}]{
\includegraphics[height=5.0cm]{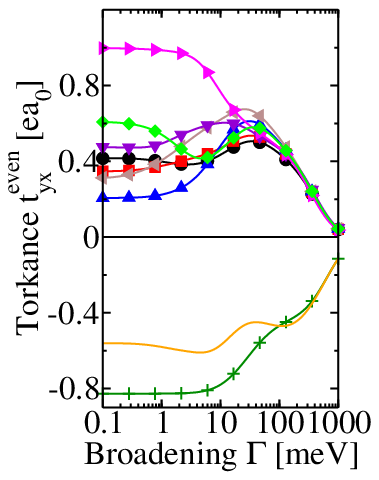}
}
\subfigure[\label{fig_effective_fields_odd}]{
\includegraphics[height=5.0cm]{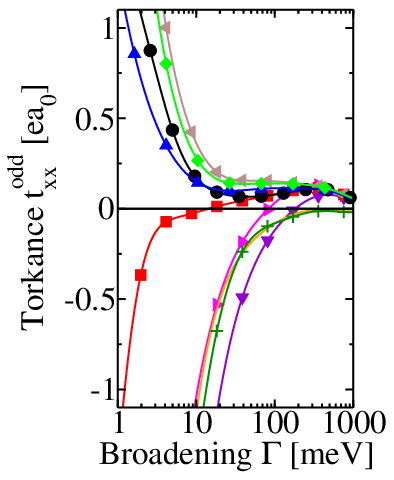}
}
\caption{\label{fig_effective_fields} 
(a) Even torkance ${\rm t}^{\rm even}_{yx}$ and 
(b) odd torkance ${\rm t}^{\rm odd}_{xx}$
in 
Co(3)/Pt(7) 
(\textbf{--\CIRCLE--}),
Co(3)/Pt(10) 
(\textbf{\textcolor[rgb]{1,0,0}{--$\blacksquare$--}}),
Co(3)/Pt(13) 
(\textbf{\textcolor[rgb]{0,1,0}{--$\Diamondblack$--}}), 
Co(3)/Pt(15) 
(\textbf{\textcolor[rgb]{0,0,1}{--\UParrow--}}),
Co(3)/Pt(20) 
(\textbf{\textcolor[rgb]{0.74,0.56,0.56}{--\LEFTarrow--}}),
Al(1)/Co(3)/Pt(10) 
(\textbf{\textcolor[rgb]{0.58,0.0,0.83}{--\DOWNarrow--}}),
O(1)/Co(3)/Pt(10)
(\textbf{\textcolor[rgb]{1,0,1}{--\RIGHTarrow--}}),
Mn(1)/W(9) 
(\textbf{\textcolor[rgb]{0,0.55,0}{--+--}})
and
Mn(1)/W(15) 
(\textbf{\textcolor[rgb]{1.0,0.65,0.0}{-----}})
for $\vn{M}$ in $z$ direction. 
The product of elementary positive charge $e$ 
and Bohr radius $a_{0}$
used as unit of torkance amounts to $ea_{0}=8.478\cdot10^{-30}$Cm.
}
\end{figure}

\section{Results}
\label{sec_results}
\subsection{Total Torkances}
In Fig.~\ref{fig_effective_fields} the torkances
${\rm t}^{\rm even}_{yx}$ and ${\rm t}^{\rm odd}_{xx}$ as obtained 
within the constant $\Gamma$ model,
Eq.~\eqref{eq_even_torque_constant_gamma} 
and Eq.~\eqref{eq_odd_torque_constant_gamma}, are plotted 
for $\vn{M}$ in $z$ direction.
For this magnetization direction, the other components of 
the torkance tensor,
${\rm t}^{\rm even}_{xx}$ and ${\rm t}^{\rm odd}_{yx}$, are zero due to the 
symmetries of the systems considered 
here (see Appendix~\ref{app_symmetry}). 
As expected from Eq.~\eqref{eq_even_torque_clean_limit},
${\rm t}^{\rm even}_{yx}$ converges to its constant
Berry curvature value in the 
limit $\Gamma\rightarrow 0$,
while ${\rm t}^{\rm odd}_{xx}$
scales like
$\Gamma^{-1}$ for small $\Gamma$,
in agreement with Eq.~\eqref{eq_odd_torque_clean_limit}.
The sign of ${\rm t}^{\rm even}_{yx}$ in all
Co/Pt(111) systems studied here is positive,
while the sign in the Mn/W(001) systems is negative. 
The positive sign of ${\rm t}^{\rm even}_{yx}$ in the Co/Pt(111)-based systems
is consistent with 
experiments.~\cite{current_induced_switching_using_spin_torque_from_spin_hall_buhrman,symmetry_spin_orbit_torques}
The negative sign of ${\rm t}^{\rm even}_{yx}$ in
the Mn/W(001) systems agrees with experiments 
on a different W-based magnetic bilayer system,
namely CoFeB/W.~\cite{stt_devices_giant_she_tungsten}
The calculated even torkances ${\rm t}^{\rm even}_{yx}$ in the Mn/W(001) systems
agree in order of magnitude to those in the Co/Pt(111)-based systems.
For broadening $\Gamma<$~30~meV the largest 
odd torkances ${\rm t}^{\rm odd}_{xx}$ among the systems studied here
are found for the Mn/W(001) systems, Al(1)/Co(3)/Pt(10) 
and O(1)/Co(3)/Pt(10).

Different heavy-metal layer thicknesses and cappings result in
differences in the local electronic structure in the magnetic layer
and at the interface between the heavy metal and the magnet.
These differences are smeared out when the broadening $\Gamma$ is large.
Therefore, both ${\rm t}^{\rm even}_{yx}$ 
and ${\rm t}^{\rm odd}_{xx}$ become
approximately independent of heavy-metal layer thickness and 
capping at large broadening $\Gamma$
in both the Mn/W(001) and the Co/Pt(111)-based systems,
while they vary substantially with layer thickness and capping 
at small $\Gamma$. 

The electrical resistivity of pure bulk Pt 
as measured experimentally at room temperature amounts to
10.6~$\mu\Omega{\rm cm}$.
If we set $\Gamma=25\,{\rm meV}$ 
the resistivities
(see also Eq.~\eqref{eq_kubo_linear_response_conductivity}) obtained
within the constant $\Gamma$ model 
amount to 
$\rho_{xx}=9.52$~$\mu\Omega{\rm cm}$ (Co(3)/Pt(7)),
$\rho_{xx}=8.85$~$\mu\Omega{\rm cm}$ (Co(3)/Pt(10)),
$\rho_{xx}=8.40$~$\mu\Omega{\rm cm}$ (Co(3)/Pt(13)),
$\rho_{xx}=8.40$~$\mu\Omega{\rm cm}$ (Co(3)/Pt(15)),
$\rho_{xx}=7.94$~$\mu\Omega{\rm cm}$ (Co(3)/Pt(20)),
$\rho_{xx}=9.90$~$\mu\Omega{\rm cm}$ (Al(1)/Co(3)/Pt(10)) and
$\rho_{xx}=10.65$~$\mu\Omega{\rm cm}$ (O(1)/Co(3)/Pt(10)).
Thus, the constant $\Gamma$ model reproduces roughly the
electrical resistivity if we set the broadening to $25\,{\rm meV}$. 
In order to estimate the torkance in the Co/Pt systems
at room temperature within the constant $\Gamma$ model
we therefore use $\Gamma=25\,{\rm meV}$.
Fig.~1(a) shows that the deviation of ${\rm t}^{\rm even}_{yx}$ from its
$\Gamma\rightarrow 0$ limit is important at $\Gamma=25\,{\rm meV}$ 
for most of the systems studied here.
In the Co/Pt(111) bilayer systems, ${\rm t}^{\rm even}_{yx}$ 
increases with Pt thickness 
from 0.5~$ea_{0}$ in Co(3)/Pt(7) to 0.68~$ea_{0}$ in Co(3)/Pt(20) 
at $\Gamma=25\,{\rm meV}$. 
Addition of an Al monolayer
increases ${\rm t}^{\rm even}_{yx}$ from 0.53~$ea_{0}$ in Co(3)/Pt(10) 
to 0.58~$ea_{0}$. Likewise,
O(1)/Co(3)/Pt(10) has a higher ${\rm t}^{\rm even}_{yx}$ of 
0.62~$ea_{0}$ in comparison to Co(3)/Pt(10).

Experimentally, the SOT is often quantified in terms of the
equivalent Oersted magnetic field that one would need to
apply in order to produce the same torque
on the magnetization like the 
SOT.~\cite{CoPtAlO_spin_torque_rashba_Gambardella,
CoPtAlO_perpendicular_switching_Gambardella,
tilting_spin_orientation_pi,
layer_thickness_dependence_current_induced_effective_field_TaCoFeB_Hayashi,
current_induced_effective_field_TaCoFeB_Suzuki} 
For a given torque $\vn{T}$
this magnetic field 
is $\vn{B}=(\vn{T}\times \hat{\vn{M}})/\mu_{\rm S}$, where
$\mu_{\rm S}$ is the total spin magnetic moment in the unit cell.
For this reason we discuss the torkance per spin magnetic 
moment, which amounts 
to ${\rm t}^{\rm even}_{yx}/\mu_{\rm S}=0.0141$~mTcm/V 
in O(1)/Co(3)/Pt(10)
(10 atomic layers of Pt are 2.3nm thick) in  
good agreement to the experimental result of 0.0139~mTcm/V
in AlO$_{x}$(2nm)/Co(0.6nm)/Pt(3nm) 
trilayers~\cite{compare_to_experiment}.  
However, in Co(3)/Pt(13) (13 atomic layers of Pt are 3 nm thick) 
the torkance per spin magnetic 
moment ${\rm t}^{\rm even}_{yx}/\mu_{\rm S}$
amounts to only 0.0091 mTcm/V.
While ${\rm t}^{\rm even}_{yx}$=0.58~$ea_{0}$ in Co(3)/Pt(13) is smaller 
than ${\rm t}^{\rm even}_{yx}$ in O(1)/Co(3)/Pt(10) by only 6\%,  
the spin magnetic moment is reduced from $\mu_{\rm S}=5.78~\mu_{\rm B}$ 
in Co(3)/Pt(13) 
to $\mu_{\rm S}=4.02~\mu_{\rm B}$ in O(1)/Co(3)/Pt(10)
due to the oxide layer. Thus, the 
increased ${\rm t}^{\rm even}_{yx}/\mu_{\rm S}$ in O(1)/Co(3)/Pt(10) 
can be attributed mainly to the
reduction of $\mu_{\rm S}$.

We now turn to the discussion of ${\rm t}^{\rm odd}_{xx}$
in the Co/Pt(111) systems. The dependence 
on Pt thickness and capping at small values of
the broadening $\Gamma$ is overall stronger than in the 
case of ${\rm t}^{\rm even}_{yx}$.
At $\Gamma=25\,{\rm meV}$ deposition of Al increases
${\rm t}^{\rm odd}_{xx}$ in magnitude from 0.025~$ea_0$ in Co(3)/Pt(10) 
to -0.835~$ea_0$, i.e., by roughly a factor of 30.
Similarly, capping by an O layer increases
${\rm t}^{\rm odd}_{xx}$ in magnitude to \mbox{-0.372}~$ea_0$.
At larger values 
of $\Gamma$, ${\rm t}^{\rm odd}_{xx}$ undergoes sign 
changes in Co(3)/Pt(10), O(1)/Co(3)/Pt(10) 
and Al(1)/Co(3)/Pt(10).
The origin of both the strong variation with capping and the 
sign changes lies in the complexity of the interfacial
spin-orbit coupling in realistic materials, where the sign of the effective 
Rashba parameter varies between different electronic 
bands leading to partial cancellation of contributions from bands with
different effective Rashba parameter.~\cite{CoPt_Haney_Stiles}  
The torkance per spin magnetic 
moment ${\rm t}^{\rm odd}_{xx}/\mu_{\rm S}=-0.0085$~mTcm/V in 
O(1)/Co(3)/Pt(10) 
for $\Gamma=25\,{\rm meV}$
is in good agreement to the experimental value 
of -0.0089~mTcm/V~\cite{compare_to_experiment} 
in non-annealed AlO$_{x}$(2nm)/Co(0.6nm)/Pt(3nm) trilayers. 
In the uncapped Co/Pt(111) systems the magnitude 
of ${\rm t}^{\rm odd}_{xx}/\mu_{\rm S}$ at $\Gamma=25\,{\rm meV}$ is
smaller and the sign is opposite to experiment. 
According to our calculations, the oxidation of Co by the 
deposition of the
AlO$_{x}$ layer is thus crucial
to obtain ${\rm t}^{\rm odd}_{xx}/\mu_{\rm S}$ as measured 
experimentally in AlO$_{x}$(2nm)/Co(0.6nm)/Pt(3nm).

W has been grown with resistivities as large as 80~$\mu\Omega{\rm cm }$
and 260~$\mu\Omega{\rm cm }$ in SOT experiments 
on CoFeB/W.~\cite{stt_devices_giant_she_tungsten}
For the high resistivity phase of W,
large SHE-angles have been reported.
Therefore, we discuss the torkances in Mn/W(001) 
at a broadening of $\Gamma=$~100~meV. At this value of broadening
we obtain within the constant $\Gamma$ model resistivities 
of $\rho_{xx}=69.4$~$\mu\Omega{\rm cm }$ and  
of $\rho_{xx}=60.2$~$\mu\Omega{\rm cm }$ in Mn(1)/W(9) 
and Mn(1)/W(15), respectively.
The corresponding torkances are
${\rm t}^{\rm even}_{yx}$~=~-0.47~$ea_0$
and
${\rm t}^{\rm odd}_{xx}$~=~\mbox{-0.082}~$ea_0$
in Mn(1)/W(9)
and
${\rm t}^{\rm even}_{yx}$~=~\mbox{-0.47}~$ea_0$
and
${\rm t}^{\rm odd}_{xx}$~=~\mbox{-0.085}~$ea_0$
in Mn(1)/W(15).
As torkances per spin magnetic moment we obtain
${\rm t}^{\rm even}_{yx}/\mu_{\rm S}$~=~\mbox{-0.131}~mTcm/V
and
${\rm t}^{\rm odd}_{xx}/\mu_{\rm S}$~=~-0.0229~mTcm/V
in Mn(1)/W(9)
and
${\rm t}^{\rm even}_{yx}/\mu_{\rm S}$~=~-0.133~mTcm/V
and
${\rm t}^{\rm odd}_{xx}/\mu_{\rm S}$~=~-0.0241~mTcm/V
in Mn(1)/W(15).
Sign and order of magnitude of ${\rm t}^{\rm even}_{yx}$
agree to the experiment on CoFeB/W.~\cite{stt_devices_giant_she_tungsten}

\subsection{Atom-resolved torkances and spin-flux coefficients}
In order to shed light on the mechanisms underlying the 
SOTs in magnetic bi- and trilayer films, we introduce the atom resolved
torkance ${\rm t}_{ij\alpha}$, which we define
by replacing $\vht{\mathcal{T}}$ in 
Eq.~\eqref{eq_kubo_linear_response}
by the operator $\vht{\mathcal{T}}_{\!\!\alpha}$, whose matrix elements are
\bege\label{eq_atom_resolved_torque_operator}
\langle\psi^{\phantom{\dagger}}_{\vn{k}n}
|
\vht{\mathcal{T}}^{\phantom{\dagger}}_{\!\!\alpha}
|
\psi^{\phantom{\dagger}}_{\vn{k}m}\rangle
=-\mu^{\phantom{\dagger}}_{\rm B}\!\!\!\!\int\limits_{\rm MT_{\alpha}}\!\!\!\!d^3r\,
\psi^{\dagger}_{\vn{k}n}(\vn{r})
\vht{\sigma}\!\times\! \vn{\Bxc}^{\rm xc}(\vn{r})
\psi^{\phantom{\dagger}}_{\vn{k}m}(\vn{r}),
\ee
where the volume integration is restricted to the 
muffin-tin sphere of the $\alpha$-th atom, which is 
denoted by MT$_{\alpha}$.
In contrast to $\vht{\mathcal{T}}$, which measures the total 
torque acting 
on the magnetization within one
unit cell, $\vht{\mathcal{T}}_{\!\!\alpha}$ probes the
torque on the spin magnetic moment
of atom $\alpha$.
Since $\vn{\Bxc}^{\rm xc}(\vn{r})$ is much larger 
inside the MT spheres than in
between them, the sum of the atom resolved torkances 
approximately yields the 
total torkance, i.e., 
${\rm t}_{ij}\approx\sum_{\alpha}{\rm t}_{ij\alpha}$.
Consequently,
the torkance on the magnetization 
in the interstitial region (INT) between the MT spheres
is small: $|{\rm t}_{ij{\rm INT}}|=|{\rm t}_{ij}-\sum_{\alpha}{\rm t}_{ij\alpha}|\ll|{\rm t}_{ij}|$. 

Additionally, we introduce the 
linear-response coefficients ${\rm q}_{ij\alpha}$
of the flux of spin angular 
momentum~\cite{current_driven_magnetization_dynamics_helical_spin_density_waves} 
into the 
MT sphere of atom $\alpha$. We define ${\rm q}_{ij\alpha}$ by 
replacing $\mathcal{T}_{i}$ in 
Eq.~\eqref{eq_kubo_linear_response} 
by the
operator $\mathcal{Q}_{i \alpha}$, the matrix elements of which are given by
\bege \label{eq_define_spin_flux_operator}
\begin{aligned}
&\langle\psi^{\phantom{\dagger}}_{\vn{k}n}|\mathcal{Q}^{\phantom{\dagger}}_{i \alpha}|\psi^{\phantom{\dagger}}_{\vn{k}m}\rangle
=
-\frac{\mu_{\rm B}\hbar}{2ie}\int\limits_{{\rm S}_{\alpha}}d\, \vn{S}\\
&\cdot\left[\psi^{\dagger}_{\vn{k}n}(\vn{r})\sigma^{\phantom{\dagger}}_{i}\vn{\nabla}\psi^{\phantom{\dagger}}_{\vn{k}m}(\vn{r})
-\vn{\nabla}\psi^{\dagger}_{\vn{k}n}(\vn{r})\sigma^{\phantom{\dagger}}_{i}\psi^{\phantom{\dagger}}_{\vn{k}m}(\vn{r})\right]
,
\end{aligned}
\ee 
where the integration is performed over the surface ${\rm S}_{\alpha}$ 
of the MT sphere of atom $\alpha$.
In the presence of SOI ${\rm q}_{ij\alpha}$
generally differs from ${\rm t}_{ij\alpha}$,
because the spin current flux can be transferred both to the
magnetization and to the
lattice and because SOI generates additional torques
that are not based on a
spin current 
flux~\cite{current_induced_torques_presence_soc,manchon_zhang_2009,torque_macdonald}.
However, SOI is negligible in the interstitial region and the spin 
angular momentum flux into the interstitial
region can therefore only be transferred to the interstitial 
magnetization 
and not to the lattice. 
Thus, we have to very good approximation
${\rm t}_{ij{\rm INT}}\approx{\rm q}_{ij{\rm INT}}$.
This flux into the interstitial region is equal to the negative 
sum of the fluxes into the
MT-spheres, i.e., ${\rm q}_{ij{\rm INT}}=-\sum_{\alpha}{\rm q}_{ij\alpha}$.
Since we argued above that $|{\rm t}_{ij{\rm INT}}|\ll|{\rm t}_{ij}|$, 
the sum of the fluxes
${\rm q}_{ij\alpha}$ is likewise small, 
i.e., $|\sum_{\alpha}{\rm q}_{ij\alpha}|\ll |{\rm t}_{ij}|$. 
Additionally, due to translational
invariance in $x$ and $y$ directions, sizable contributions to 
${\rm q}_{ij\alpha}$ can only originate from spin currents flowing
in $z$ direction. Thus, the spin fluxes ${\rm q}_{ij\alpha}$ 
indicate by how much the non-equilibrium spin current flowing in $z$
direction is modified as it traverses the $\alpha$-th atomic layer.

\begin{figure}
\flushleft
\includegraphics[width=8.95cm]{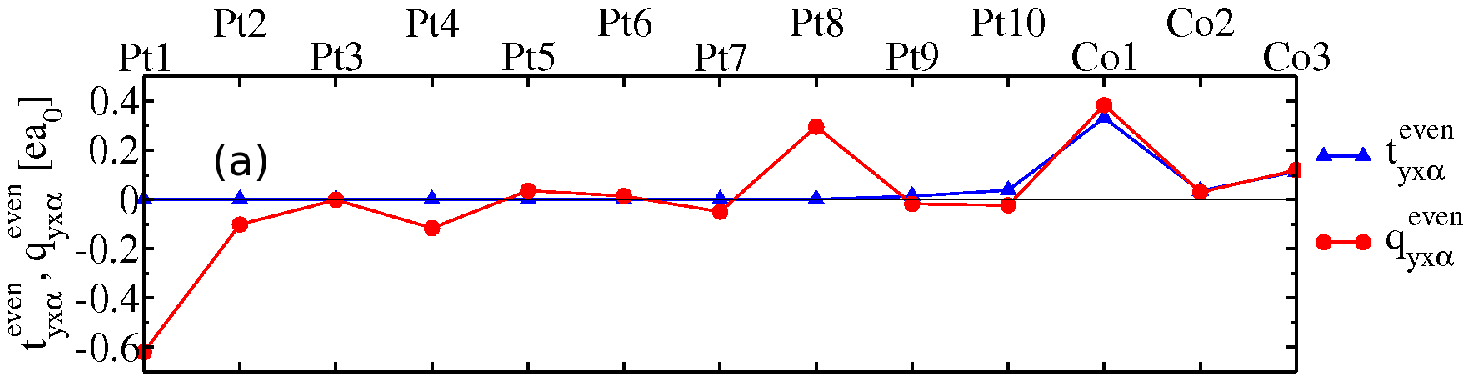}\\
\includegraphics[width=8.5cm]{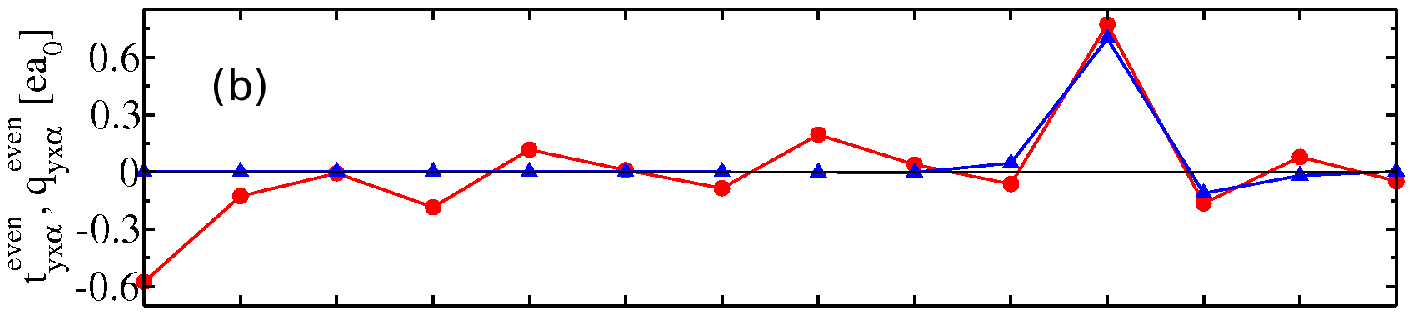}\\
\includegraphics[width=8.5cm]{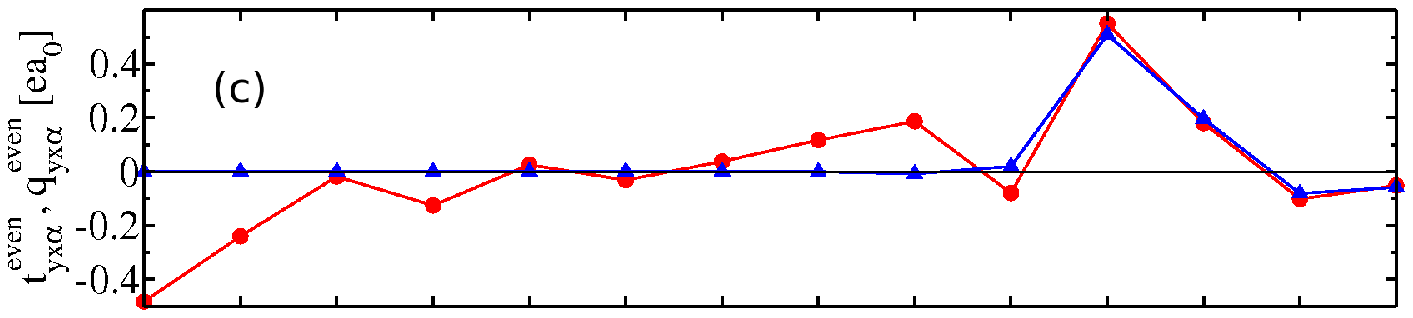}\\
\includegraphics[width=8.92cm]{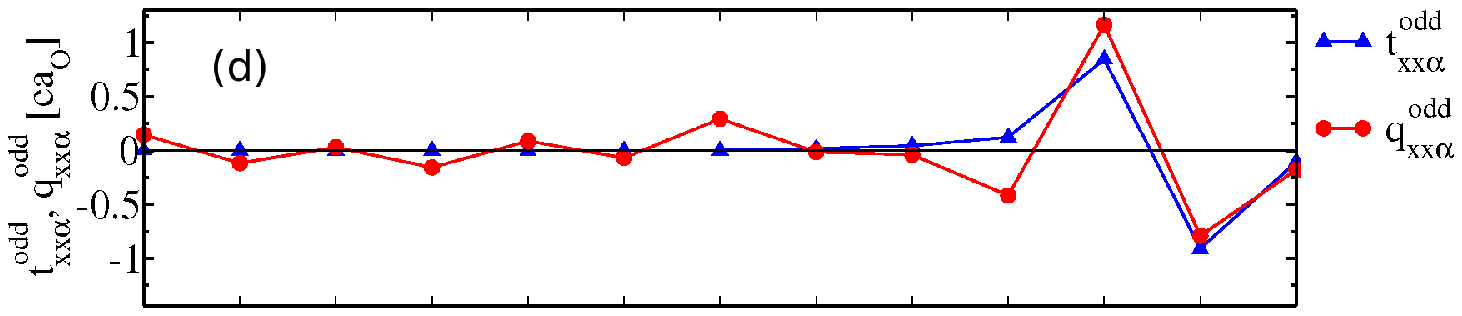}\\
\includegraphics[width=8.5cm]{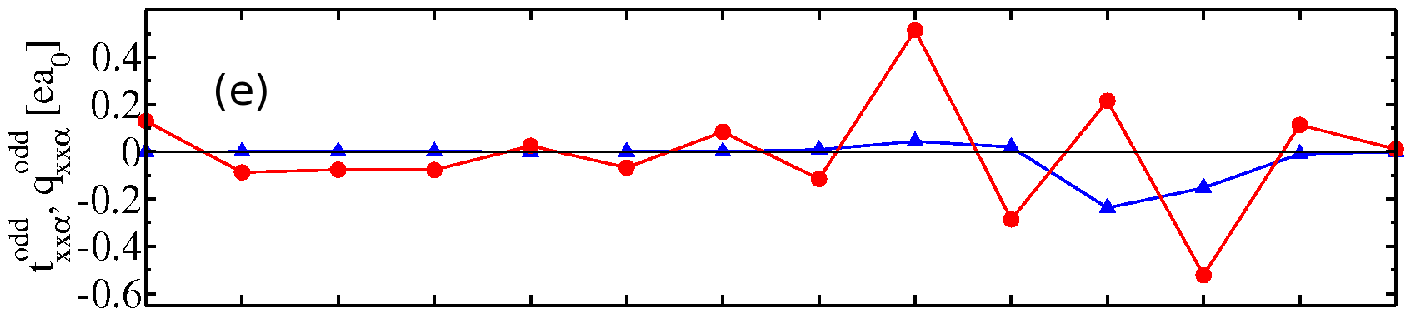}\\
\includegraphics[width=8.62cm]{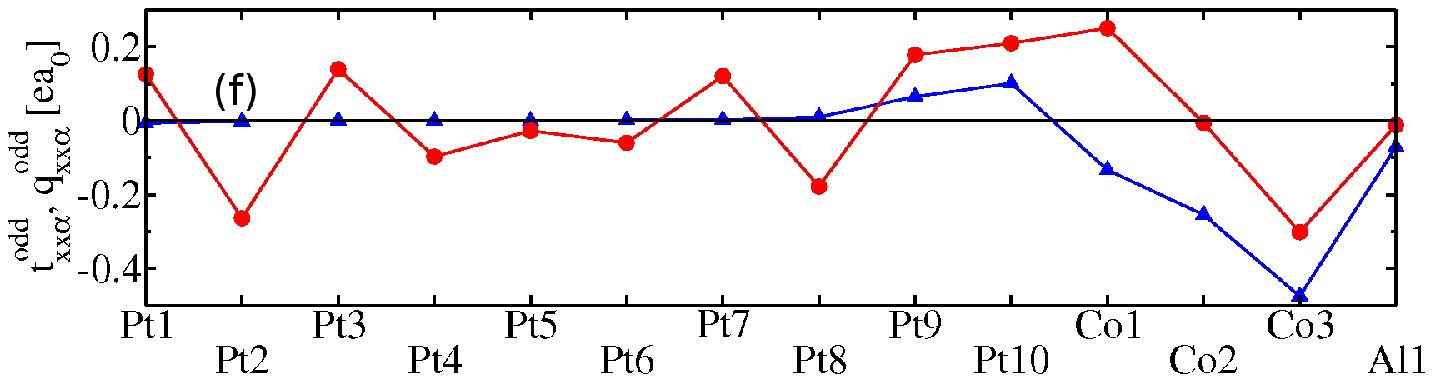}
\caption{\label{fig_torques_vs_spincurrents}
Atom resolved torkance ${\rm t}^{\rm even}_{yx\alpha}$ (triangles) 
and atom resolved spin-flux 
coefficient ${\rm q}^{\rm even}_{yx\alpha}$ (circles) in 
(a) Co(3)/Pt(10), 
(b) O(1)/Co(3)/Pt(10) 
and (c) Al(1)/Co(3)/Pt(10).  
Atom resolved torkance ${\rm t}^{\rm odd}_{xx\alpha}$ (triangles) 
and atom resolved spin-flux 
coefficient ${\rm q}^{\rm odd}_{xx\alpha}$ (circles) in 
(d) Co(3)/Pt(10), 
(e) O(1)/Co(3)/Pt(10) 
and (f) Al(1)/Co(3)/Pt(10). 
The broadening was set to $\Gamma=25$~meV. Lines serve as guides to the eye.}
\end{figure}

In Fig.~\ref{fig_torques_vs_spincurrents}(a)-(c) the atom resolved 
torkances ${\rm t}^{\rm even}_{yx\alpha}$
and spin-flux coefficients ${\rm q}^{\rm even}_{yx\alpha}$ are shown for
the systems Co(3)/Pt(10), O(1)/Co(3)/Pt(10) and Al(1)/Co(3)/Pt(10) 
at $\Gamma=25$~meV. 
${\rm t}^{\rm even}_{yx\alpha}$ and ${\rm q}^{\rm even}_{yx\alpha}$ are obtained
by replacing $\vht{\mathcal{T}}$ in 
Eq.~\eqref{eq_even_torque_constant_gamma}
by $\vht{\mathcal{T}}_{\!\!\alpha}$
and $\vht{\mathcal{Q}}_{\alpha}$, respectively.
${\rm t}^{\rm even}_{yx\alpha}$ 
(shown as triangles) 
is strongest at Co1 and negligibly small at Pt atoms.
The difference between 
${\rm t}^{\rm even}_{yx\alpha}$ 
and
${\rm q}^{\rm even}_{yx\alpha}$ 
(shown as circles)
is insignificant at all three Co atoms.
Thus, in the systems considered here, 
the even torkance ${\rm t}^{\rm even}_{yx\alpha}$ arises 
from the spin flux into the Co-layer. In this regard, 
the even torque on the Co magnetization 
resembles the spin-transfer torque (STT): In the case of STT in
spin valves spin current is generated in one magnetic layer
that acts as a polarizer and transferred to the magnetization of a 
second magnetic layer that acts as an analyzer.~\cite{stiles_miltat_stt} 
In the Co/Pt(111) systems considered here,
spin angular momentum flows from the nonmagnetic Pt layer
into the Co layer, where it produces a torque on the magnetization.

The coefficient of spin-flux into the Co-layer, i.e.,
${\rm q}^{\rm even}_{yx{\rm Co1}}+{\rm q}^{\rm even}_{yx{\rm Co2}}+
{\rm q}^{\rm even}_{yx{\rm Co3}}$, is positive. This implies
a spin current with spin polarization along the $+y$ direction 
flowing in $+z$ direction for an electric field along the $+x$ direction.
This sign of spin current agrees to the one of the intrinsic SHE
of bulk 
Pt.~\cite{current_induced_switching_using_spin_torque_from_spin_hall_buhrman,
Guo_Murakami_Chen_Nagaosa}
For $\alpha$ in the Pt layer ${\rm q}^{\rm even}_{yx\alpha}$
tends to be relatively small except for $\alpha$=Pt1.
The coefficient ${\rm q}^{\rm even}_{yx{\rm Pt1}}$ is negative
and thus opposite in sign to the spin-flux coefficient on Co1.
This negative spin-flux into Pt1 arises from the absorption of
spin current with spin polarization along the $-y$ direction 
flowing in $-z$ direction, which is
equivalent
to spin current with spin polarization along the $+y$ direction 
flowing in $+z$ direction. 
Hence, both the negative ${\rm q}^{\rm even}_{yx{\rm Pt1}}$
and the positive ${\rm q}^{\rm even}_{yx{\rm Co1}}$ are consistent with
a spin current in Pt that is characterized by a
spin polarization along $+y$ and 
flows in $+z$ direction for electric field applied 
along the $+x$ direction.
This spin current is absorbed efficiently by the magnetic Co1 atoms
as well as by the nonmagnetic Pt1 atoms. The
spin current absorbed at the Co1 atoms is transferred to 
the Co magnetization, while the spin current absorbed at the Pt1 atoms
is transferred to the lattice via the spin-orbit interaction.

The situation is rather different for ${\rm t}^{\rm odd}_{xx\alpha}$ 
and ${\rm q}^{\rm odd}_{xx\alpha}$ shown 
in Fig.~\ref{fig_torques_vs_spincurrents}(d)-(f). The cases 
of O(1)/Co(3)/Pt(10) and Al(1)/Co(3)/Pt(10) show clearly that 
in general
the torkance on the Co moments differs significantly from the 
spin-flux coefficient. 
These differences between ${\rm t}^{\rm odd}_{xx\alpha}$ 
and ${\rm q}^{\rm odd}_{xx\alpha}$ for $\alpha$=Co1, Co2, Co3 
result from the SOI on the
Co atoms, which allows on the one hand the transfer of spin angular 
momentum flux to 
the lattice and on the other hand the generation of torques on the 
magnetization that
are not related to a spin angular momentum flux.
In the cases of Co(3)/Pt(10) and O(1)/Co(3)/Pt(10) the coefficients
${\rm q}^{\rm odd}_{xx\alpha}$ are very small for $\alpha$=Pt1 through
$\alpha$=Pt8, in contrast to ${\rm q}^{\rm even}_{yx\alpha}$ discussed
above, where in particular ${\rm q}^{\rm even}_{yx{\rm Pt1}}$ is found to
be sizable. This means that the spin fluxes that contribute to
the odd torques in Co(3)/Pt(10) 
and O(1)/Co(3)/Pt(10) originate only close to the 
interface ($\alpha$=Pt9 and $\alpha$=Pt10). Thus, the interfacial spin-orbit
coupling rather than the bulk Pt spin-orbit coupling contributes to
the odd torque in Co(3)/Pt(10) 
and O(1)/Co(3)/Pt(10). In Al(1)/Co(3)/Pt(10) ${\rm q}^{\rm odd}_{xx\alpha}$ 
is sizable for $\alpha$=Pt1 through $\alpha$=Pt3, 
but ${\rm q}^{\rm odd}_{xx\alpha}$ oscillates in this region such that the sum of
spin-fluxes from the region $\alpha$=Pt1 through $\alpha$=Pt3 is negligible.
Hence, the spin-flux contribution to the odd torque in Al(1)/Co(3)/Pt(10)
originates also in this case from the Co/Pt interface region.

Thus, the comparison of Fig.~\ref{fig_torques_vs_spincurrents}(a)-(c) on the
one hand to Fig.~\ref{fig_torques_vs_spincurrents}(d)-(f) on the other hand
corroborates the picture that the even torque in the Co/Pt system is
associated with a spin current in Pt arising mainly
from the bulk Pt spin orbit
interaction and flowing into the Co layer, while the odd torque is associated
with interfacial spin-orbit coupling. While the electronic structure in the
thin Pt layer differs from the one in bulk Pt, the bulk Pt SHE is 
predictive in both sign and order of magnitude
for the spin current causing the even 
torque.~\cite{current_induced_switching_using_spin_torque_from_spin_hall_buhrman}  

\begin{figure}
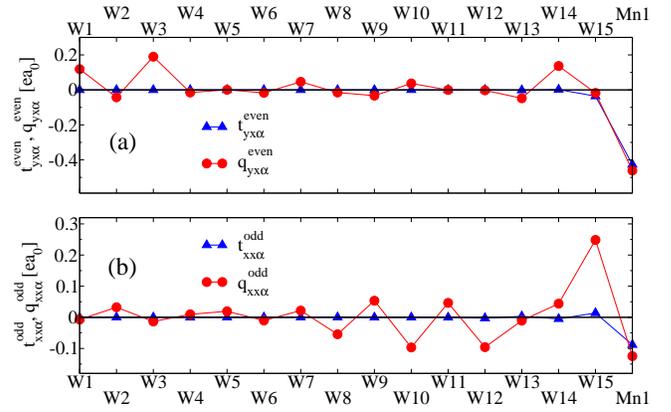

\includegraphics[width=8.5cm]{Mn1W15_atomresolved_torques_100meV.eps}\\
\vspace{0.2cm}
\includegraphics[width=8.5cm]{Mn1W15_atomresolved_odd_torques_100meV.eps}\\
\caption{\label{fig_torques_vs_spincurrents_Mn1W15}
Atom resolved torkances (triangles) and
spin-flux coefficients (circles) in Mn(1)/W(15) at
$\Gamma=$100~meV. (a) ${\rm t}^{\rm even}_{yx\alpha}$ 
and ${\rm q}^{\rm even}_{yx\alpha}$.
(b) ${\rm t}^{\rm odd}_{xx\alpha}$ 
and ${\rm q}^{\rm odd}_{xx\alpha}$.
}
\end{figure}

Fig.~\ref{fig_torques_vs_spincurrents_Mn1W15} shows atom resolved
torkances and spin-flux coefficients in the Mn(1)/W(15) system. In the 
case of ${\rm t}^{\rm even}_{yx\alpha}$ 
and ${\rm q}^{\rm even}_{yx\alpha}$ shown 
in Fig.~\ref{fig_torques_vs_spincurrents_Mn1W15}(a)
we can identify a middle region ($\alpha$=W4 through $\alpha$=W13) where
${\rm q}^{\rm even}_{yx\alpha}$ is small. Large spin-fluxes exist for the
atoms W1 and W3 as well as for Mn1. 
The sign of the spin-fluxes into W1 and W3 
is opposite to the one for Mn1. Thus, the sign of these 
spin fluxes is consistent
with a spin current with spin polarization along $-y$ 
flowing in $+z$ direction if the electric field is applied
along $+x$ direction. This spin current generates a torque 
on the magnetization of the Mn1 layer, whereby it is absorbed.
Since the difference between ${\rm t}^{\rm even}_{yx{\rm Mn1}}$ 
and ${\rm q}^{\rm even}_{yx{\rm Mn1}}$ is small, the torque on Mn1 arises
dominantly from the spin-flux into the Mn1 sphere.
The sign of the spin current generated in W is opposite to the one in Pt and
consistent with the intrinsic SHE in 
bulk W.~\cite{prb_she_tanaka,stt_devices_giant_she_tungsten}
On the other hand,  
${\rm q}^{\rm odd}_{xx\alpha}$ shown 
in Fig.~\ref{fig_torques_vs_spincurrents_Mn1W15}(b)
is negligible in the region $\alpha$=W1 through $\alpha$=W7.
The exchange of angular momentum between W and Mn that contributes
to the odd torque is thus restricted to
the interfacial region and can be attributed
to the interfacial spin orbit interaction.
Since the difference between ${\rm t}^{\rm odd}_{xx{\rm Mn1}}$ 
and ${\rm q}^{\rm odd}_{xx{\rm Mn1}}$ is small, also the odd torque arises
mostly from a spin-flux in this case. In summary, the qualitative
behavior of the atom resolved torkances
and spin-flux coefficients in the Mn(1)/W(15) system 
resembles the one in the Co(3)/Pt(10) system.

\begin{figure}
\includegraphics[width=8.5cm]{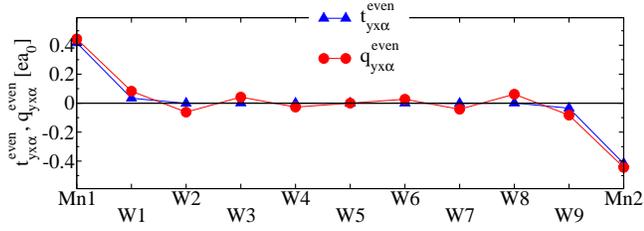}\\
\caption{\label{fig_even_torques_vs_spincurrents_WMnW}
Atom resolved torkance ${\rm t}^{\rm even}_{yx\alpha}$ (triangles)
and atom resolved spin-flux 
coefficient ${\rm q}^{\rm even}_{yx\alpha}$ (circles) in
Mn(1)/W(9)/Mn(1) at $\Gamma$=100~meV.}
\end{figure}

The large values of ${\rm q}^{\rm even}_{yx{\rm Pt1}}$ 
and ${\rm q}^{\rm even}_{yx{\rm W1}}$ in 
Fig.~\ref{fig_torques_vs_spincurrents}(a)-(c)
and
Fig.~\ref{fig_torques_vs_spincurrents_Mn1W15}(a), 
respectively, pose the question
to which extent an additional substrate below the heavy metal 
layer might influence
the even torque, in particular if the heavy metal layer is thin. 
In order to address this question, we computed atom-resolved
torkances and spin-flux coefficients in Mn(1)/W(9)/Mn(1), which we show in
Fig.~\ref{fig_even_torques_vs_spincurrents_WMnW}. The magnetizations in 
the two Mn layers, Mn1 and Mn2, are chosen to be
parallel to each other and along the $z$ direction. 
Comparing 
Fig.~\ref{fig_even_torques_vs_spincurrents_WMnW} to
Fig.~\ref{fig_torques_vs_spincurrents_Mn1W15}(a) we find that 
in the Mn(1)/W(9)/Mn(1) system the
additional magnetic layer (Mn1) absorbs the spin current with 
spin-polarization along $+y$, which is flowing in $-z$ direction, while
in the Mn(1)/W(15) system it is absorbed by W1 and W3.
Both absorption mechanisms are sufficiently efficient 
to prevent transformation
of significant portions of the spin current 
flowing in $-z$ direction into a spin current flowing
in $+z$ direction by reflection at the boundary of the system. 
Consequently, the  
resulting torkance ${\rm t}^{\rm even}_{yx{\rm Mn2}}$ in the 
Mn(1)/W(9)/Mn(1) system agrees well with the 
torkance ${\rm t}^{\rm even}_{yx{\rm Mn1}}$ in the Mn(1)/W(15) system. 

\begin{figure}
\includegraphics[width=9cm]{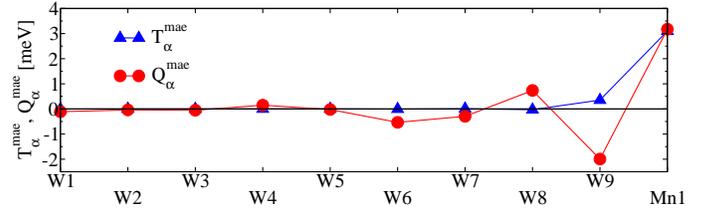}
\caption{\label{fig_mae_torque_MnW}
Atom resolved torques (triangles) and
spin-fluxes (circles) in Mn(1)/W(9)
without applied electric field 
when the magnetization is rotated away from the
easy axis by 30$^{\circ}$.
}
\end{figure}

The torkance ${\rm t}_{ij}^{\rm odd}$ as given by 
Eq.~\eqref{eq_odd_torque_clean_limit} can be interpreted
as correction to the magnetic anisotropy due to the
nonequilibrium distribution of electrons.~\cite{torque_macdonald}
Without an applied electric field the torque due to magnetic anisotropy
is given by~\cite{mae_torque_method}
\bege\label{eq_mae_torque}
\vn{T}^{\rm mae}
=-\frac{1}{\mathcal{N}}\sum_{\vn{k}n}
f^{\phantom{R}}_{\vn{k}n}
\langle
\psi^{\phantom{R}}_{\vn{k}n}
|
\vht{\mathcal{T}}
|
\psi^{\phantom{R}}_{\vn{k}n}
\rangle
,
\ee
where $f^{\phantom{R}}_{\vn{k}n}$ is 1 for occupied states and zero
otherwise. Within the relaxation time approximation,
an applied electric field $\vn{E}$ changes the 
occupancies $f^{\phantom{R}}_{\vn{k}n}$ by
$\delta f^{\phantom{R}}_{\vn{k}n}=-e\tau
\langle\psi^{\phantom{R}}_{\vn{k}n}
|
\vn{v}
|
\psi^{\phantom{R}}_{\vn{k}n}
\rangle
\cdot\vn{E}\,
\delta(\mathcal{E}^{\phantom{R}}_{\rm F}
-
\mathcal{E}^{\phantom{R}}_{\vn{k}n})$, which modifies
the torque by
\bege\label{eq_neq_torque}
-\frac{1}{\mathcal{N}}\sum_{\vn{k}n}
\delta f^{\phantom{R}}_{\vn{k}n}
\langle
\psi^{\phantom{R}}_{\vn{k}n}
|
\mathcal{T}_{i}
|
\psi^{\phantom{R}}_{\vn{k}n}
\rangle
=\sum_j
{\rm t}^{\rm odd}_{ij}{\rm E}_{j},
\ee
where ${\rm E}_{j}$ are the cartesian components of the applied
electric field and ${\rm t}^{\rm odd}_{ij}$ is given 
by Eq.~\eqref{eq_odd_torque_clean_limit}.
As discussed above, part of the odd torque is mediated by
spin currents. In order to demonstrate that also the magnetic
anisotropy torque, Eq.~\eqref{eq_mae_torque}, can
contain important contributions from
spin currents in bilayer systems, we investigate
the atom-resolved torques $\vn{T}_{\alpha}^{\rm mae}$ as well as the atom-resolved
spin-fluxes $\vn{Q}_{\alpha}^{\rm mae}$, which are obtained from
Eq.~\eqref{eq_mae_torque} by replacing the torque operator
$\vht{\mathcal{T}}$
by the atom resolved torque operator $\vht{\mathcal{T}}_{\!\!\alpha}$,
Eq.~\eqref{eq_atom_resolved_torque_operator},
and the atom resolved spin flux $\vht{\mathcal{Q}}_{\alpha}$,
Eq.~\eqref{eq_define_spin_flux_operator}, respectively. 
Fig.~\ref{fig_mae_torque_MnW} shows $\vn{T}_{\alpha}^{\rm mae}$ 
and $\vn{Q}_{\alpha}^{\rm mae}$
in Mn(1)/W(9) when the magnetization is tilted
away from the $z$ axis towards the $x$ axis by 30$^{\circ}$ (for
magnetization along $z$, i.e., along the easy axis, 
both $\vn{T}_{\alpha}^{\rm mae}$ and $\vn{Q}_{\alpha}^{\rm mae}$ are zero). 
Clearly, the torque on Mn1 arises almost entirely from the spin flux.
The spin fluxes $\vn{Q}_{\alpha}^{\rm mae}$ decay rapidly with increasing
distance from the
interface and are negligible on W1 through W5. This behavior resembles
the one of the spin flux coefficients ${\rm q}^{\rm odd}_{xx\alpha}$ shown
in Fig.~\ref{fig_torques_vs_spincurrents_Mn1W15}~(b).

The electron wave functions are 
spinors $\psi^{\phantom{R}}_{\vn{k}n}(\vn{r})
=(\psi^{\phantom{R}}_{\vn{k}n\uparrow}(\vn{r}),
\psi^{\phantom{R}}_{\vn{k}n\downarrow}(\vn{r}))^{\rm T}$, which do not
carry any charge current in $z$ direction. However the
spin-up and spin-down components of these spinors separately carry equal but
opposite charge currents in $z$ direction. Thereby a spin current
in $z$ direction is associated with each Bloch 
function $|\psi^{\phantom{R}}_{\vn{k}n}\rangle$. These spin currents 
in $z$ direction interact with the spin-orbit interaction as well as with
the exchange field in Mn1 and close to the interface. Thereby, 
they exhibit the spin fluxes $\vn{Q}_{\alpha}^{\rm mae}$ shown in 
Fig.~\ref{fig_mae_torque_MnW} and contribute
to the magnetic anisotropy. This mechanism resembles the interlayer
exchange coupling in spin-valves or tunnel junctions, which is 
mediated by spin currents 
that flow between two magnets
even in the absence of an
applied bias.~\cite{PhysRevB.39.6995,PhysRevB.79.054405} In comparison
to interlayer exchange coupling in spin valves, 
the spin-orbit interaction takes over
the role of one of the two ferromagnets.

For magnetization along $z$ individual 
states $|\psi^{\phantom{R}}_{\vn{k}n}\rangle$ in Eq.~\eqref{eq_mae_torque}
exhibit non-zero torques 
$\langle
\psi^{\phantom{R}}_{\vn{k}n}
|
\vht{\mathcal{T}}
|
\psi^{\phantom{R}}_{\vn{k}n}
\rangle$
and spin fluxes
$\langle
\psi^{\phantom{R}}_{\vn{k}n}
|
\vht{\mathcal{Q}}^{\phantom{R}}_{\alpha}
|
\psi^{\phantom{R}}_{\vn{k}n}
\rangle$.
However, the net torques and spin fluxes are zero when the Brillouin zone
summation is carried out and when magnetization is along $z$. 
For this reason we tilted the 
magnetization direction away from the easy 
axis in Fig.~\ref{fig_mae_torque_MnW}. When an electric field is applied
the states are occupied according to a nonequilibrium distribution
and the Brillouin zone summation in Eq.~\eqref{eq_neq_torque} yields 
a non-zero torque even for
magnetization along $z$. This explains the similar qualitative behaviour
of the odd spin flux coefficients ${\rm q}^{\rm odd}_{xx\alpha}$ shown
in Fig.~\ref{fig_torques_vs_spincurrents_Mn1W15}~(b) and the spin
fluxes $\vn{Q}_{\alpha}^{\rm mae}$: The spin currents that contribute 
to the odd torque
are present also without applied electric field. An
additional electric field only changes the relative weight of the spin current
associated with a given state $|\psi^{\phantom{R}}_{\vn{k}n}\rangle$ by 
changing its occupancy $f^{\phantom{R}}_{\vn{k}n}$. In contrast, the spin
current due to SHE is not present without applied electric field, which is
why the even spin flux coefficients ${\rm q}^{\rm even}_{yx\alpha}$ differ
qualitatively from ${\rm q}^{\rm odd}_{xx\alpha}$.

\section{Summary}
We performed first principles calculations of the SOTs 
in Co/Pt(111) and Mn/W(001)
within
the Kubo linear-response formalism. 
We decomposed the SOTs into their even and odd
contributions with respect to magnetization reversal,
because these even and odd parts depend differently on
disorder, which we approximated by a constant band broadening.
Moreover, in the bi- and trilayer systems considered here,
the even torque arises dominantly from bulk spin-orbit coupling
in the heavy metal layer, while the odd torque depends strongly
on the interfacial spin-orbit coupling.
Moreover, we found that the even and the
odd torque can be of similar magnitude, but that 
the odd torque can be
suppressed due to its strong dependence on details like the capping and
the heavy metal layer thickness.
While the even torque is almost
entirely explained by spin-currents originating in the heavy metal
layer and flowing into
the magnetic layer, the odd torque can contain a sizable 
contribution that does
not stem from spin transfer. The spin currents which add to the
odd torque are also present if no electric field is applied, in which case
they contribute to the magnetic anisotropy.
Our results are in satisfactory agreement with experimental measurements.

\acknowledgments
We gratefully acknowledge discussions with G.~Bihlmayer, 
P.~Gambardella, 
K.~Garello and I.~M.~Miron,
computing time on the supercomputers \mbox{JUQUEEN} 
and \mbox{JUROPA} 
at J\"ulich Supercomputing Center and funding under 
the HGF-YIG programme VH-NG-513.

\appendix
\section{Linear-Response formalism for the torkance}
\label{app_sot_formula}
We can write the torque on the magnetization within one unit cell
due to an electron in state $\psi(\vn{r})$
as
\bege
\begin{aligned}
\vn{T}[\psi]&=-\mu_{\rm B}
\int d^{3}r
\vn{\Bxc}^{\rm xc}(\vn{r})\times \psi(\vn{r})^{\dagger}\vht{\sigma}\psi(\vn{r})\\
&=\mu_{\rm B}\int d^{3}r
\psi(\vn{r})^{\dagger}\vht{\sigma}\psi(\vn{r})\times \vn{\Bxc}^{\rm xc}(\vn{r})\\
&=-\int d^{3}r \psi(\vn{r})^{\dagger}\vht{\mathcal{T}}(\vn{r})\psi(\vn{r}),
\end{aligned}
\ee
where $\vht{\mathcal{T}}(\vn{r})=-\mu_{\rm B}\vht{\sigma}\times\vn{\Bxc}^{\rm xc}(\vn{r})$ is the
torque operator. 
According to Kubo linear-response theory~\cite{book_mahan}, the
torkance tensor is given by
\bege\label{eq_torkance}
{\rm t}_{ij}=-\lim\limits_{\mathcal{E}\rightarrow 0}
\left[
\frac{1}{\mathcal{E}} \,{\rm Im} \,\Pi_{ij}(\mathcal{E})
\right],
\ee
where $\Pi_{ij}(\mathcal{E})$ is $e$ times the Fourier transform of the
retarded  torque-velocity correlation function:
\bege
\Pi_{ij}(\mathcal{E})=-ie\int\limits_{0}^{\infty}dte^{\frac{i}{\hbar}\mathcal{E}t}
\langle
[\mathcal{T}_{i}(t),v_{j}(0)]_{-}
\rangle.
\ee
Following the standard recipe~\cite{book_mahan} to obtain 
retarded functions conveniently
by analytical continuation of Matsubara functions, 
it is straightforward to derive the expressions
\begin{gather}\label{eq_kubo_linear_response_app}
\begin{aligned}
{\rm t}^{\rm I(a)\phantom{I}}_{ij}\!\!\!\!=-\frac{e}{h}\int_{-\infty}^{\infty}&
d\mathcal{E}
\frac{d f(\mathcal{E})}{d \mathcal{E}}
\phantom{{\rm Re}}
{\rm Tr}
\langle\mathcal{T}_{i}
G^{\rm R}(\mathcal{E})
v_{j}
G^{\rm A}(\mathcal{E})
\rangle_{\rm c}
\\
{\rm t}^{\rm I(b)\phantom{I}}_{ij}\!\!\!\!=\phantom{-}\frac{e}{h}\int_{-\infty}^{\infty}
&d\mathcal{E}\frac{d f(\mathcal{E})}{d \mathcal{E}}
{\rm Re}
{\rm Tr}
\langle\mathcal{T}_{i}
G^{\rm R}(\mathcal{E})
v_{j}
G^{\rm R}(\mathcal{E})
\rangle_{\rm c}\phantom{(1)}
\\
{\rm t}^{\rm II\phantom{(a)}}_{ij}\!\!\!\!=\phantom{-}\frac{e}{h}\int_{-\infty}^{\infty}
&d\mathcal{E} f(\mathcal{E})
\quad\!\!
{\rm Re}{\rm Tr}\langle
\mathcal{T}_{i}G^{\rm R}(\mathcal{E})v_{j}
\frac{dG^{\rm R}(\mathcal{E})}{d\mathcal{E}}\\
 &\quad\quad\quad\quad\quad\,-
\mathcal{T}_{i}\frac{dG^{\rm R}(\mathcal{E})}{d\mathcal{E}}v_{j}G^{\rm R}(\mathcal{E})
\rangle_{\rm c},
\end{aligned}\raisetag{4\baselineskip}
\end{gather}
which give the torkance as sum of three 
terms, ${\rm t}_{ij}={\rm t}^{\rm I(a)\phantom{I}}_{ij}
+{\rm t}^{\rm I(b)\phantom{I}}_{ij}
+{\rm t}^{\rm II\phantom{(a)}}_{ij},$ 
the first two of which are Fermi surface terms, 
while the third one is a Fermi sea term.
We assume that the dominant effect of room temperature on the 
torkance is the enhancement of the band broadening
$\Gamma$. Therefore, we set the temperature of the Fermi-Dirac 
distribution function $f(\mathcal{E})$ to zero in
the calculations. Replacing in 
Eq.~\eqref{eq_kubo_linear_response_app} $f(\mathcal{E})$ by the Heaviside step
function $\theta(\mathcal{E}_{\rm F}-\mathcal{E})$ and
$\frac{d f(\mathcal{E})}{d \mathcal{E}}$ by the Dirac delta 
function $-\delta(\mathcal{E}_{\rm F}-\mathcal{E})$
leads to Eq.~\eqref{eq_kubo_linear_response} in the main text.

The result Eq.~\eqref{eq_kubo_linear_response_app} can also be obtained 
from the Bastin equation~\cite{crepieux_bruno_ahe,bastin_conductivity}
for the conductivity tensor,
\begin{gather}\label{eq_kubo_linear_response_conductivity}
\begin{aligned}
\sigma^{\rm I(a)\phantom{I}}_{ij}\!\!\!\!=-\frac{e^{2}}{hV}\int_{-\infty}^{\infty}&
d\mathcal{E}
\frac{d f(\mathcal{E})}{d \mathcal{E}}
\phantom{{\rm Re}}
{\rm Tr}
\langle v_{i}
G^{\rm R}(\mathcal{E})
v_{j}
G^{\rm A}(\mathcal{E})
\rangle_{\rm c}
\\
\sigma^{\rm I(b)\phantom{I}}_{ij}\!\!\!\!=\phantom{-}\frac{e^2}{hV}\int_{-\infty}^{\infty}
&d\mathcal{E}\frac{d f(\mathcal{E})}{d \mathcal{E}}
{\rm Re}
{\rm Tr}
\langle v_{i}
G^{\rm R}(\mathcal{E})
v_{j}
G^{\rm R}(\mathcal{E})
\rangle_{\rm c}\phantom{(1)}
\\
\sigma^{\rm II\phantom{(a)}}_{ij}\!\!\!\!=\phantom{-}\frac{e^2}{hV}\int_{-\infty}^{\infty}
&d\mathcal{E} f(\mathcal{E})
\quad\!\!
{\rm Re}{\rm Tr}\langle
v_{i}G^{\rm R}(\mathcal{E})v_{j}
\frac{dG^{\rm R}(\mathcal{E})}{d\mathcal{E}}\\
 &\quad\quad\quad\quad\quad\,-
v_{i}\frac{dG^{\rm R}(\mathcal{E})}{d\mathcal{E}}v_{j}G^{\rm R}(\mathcal{E})
\rangle_{\rm c},
\end{aligned}\raisetag{4\baselineskip}
\end{gather}
by replacing the current density operator $-ev_{i}/V$ by $-\mathcal{T}_{i}$.

Within the constant $\Gamma$ model it is convenient to use the eigenstate
representation, 
i.e., $G^{\rm R}_{\vn{k}n}(\mathcal{E})=
\hbar[\mathcal{E}-\mathcal{E}^{\phantom{R}}_{\vn{k}n}+i\Gamma]^{-1}$.
The eigenstate representation allows us 
to split Eq.~\eqref{eq_kubo_linear_response} 
into two terms. One term,
which contains only contributions of
${\rm Re}\left[ \langle \psi_{\vn{k}n}  |\mathcal{T}_{i}| \psi_{\vn{k}m}  \rangle
\langle \psi_{\vn{k}m}  |v_{j}| \psi_{\vn{k}n}  \rangle\right]$
and a second term containing only contributions of
${\rm Im}\left[ \langle \psi_{\vn{k}n}  |\mathcal{T}_{i}| \psi_{\vn{k}m}  \rangle
\langle \psi_{\vn{k}m}  |v_{j}| \psi_{\vn{k}n}  \rangle\right]$.
The first term is given by
\bege
{\rm t}^{\rm odd}_{ij}=\frac{e\hbar}{\pi\mathcal{N}}
\sum_{\vn{k}nm}
\frac{\Gamma^2
{\rm Re}
\left[ \langle \psi_{\vn{k}n}  |\mathcal{T}_{i}| \psi_{\vn{k}m}  \rangle
\langle \psi_{\vn{k}m}  |v_{j}| \psi_{\vn{k}n}  \rangle\right]
}{
\left[(\mathcal{E}^{\phantom{R}}_{\rm F}-\mathcal{E}^{\phantom{R}}_{\vn{k}n})^2+\Gamma^2\right]
\left[(\mathcal{E}^{\phantom{R}}_{\rm F}-\mathcal{E}^{\phantom{R}}_{\vn{k}m})^2+\Gamma^2\right]
}.
\ee
Only the Fermi surface terms in 
Eq.~\eqref{eq_kubo_linear_response_app}, i.e., ${\rm t}_{ij}^{\rm I(a)}$
and ${\rm t}_{ij}^{\rm I(b)}$, contribute to ${\rm t}^{\rm odd}_{ij}$.
Using the transformation properties under time reversal
\bege\label{eq_time_reversal}
\begin{aligned}
\vn{\Bxc}^{\rm xc}&\rightarrow-\vn{\Bxc}^{\rm xc}\\
\langle\psi_{\vn{k}n}|\vn{v}|\psi_{\vn{k}m}\rangle
&\rightarrow -
(\langle\psi_{\vn{k}n}|\vn{v}|\psi_{\vn{k}m}\rangle)^{*}\\
\langle\psi_{\vn{k}n}|\vn{m}|\psi_{\vn{k}m}\rangle
&\rightarrow -
(\langle\psi_{\vn{k}n}|\vn{m}|\psi_{\vn{k}m}\rangle)^{*}\\
\langle\psi_{\vn{k}n}|\vht{\mathcal{T}}|\psi_{\vn{k}m}\rangle
&\rightarrow 
(\langle\psi_{\vn{k}n}|\vht{\mathcal{T}}|\psi_{\vn{k}m}\rangle)^{*}\\
\end{aligned}
\ee
it is straightforward to show that 
the torkance component ${\rm t}^{\rm odd}_{ij}$ is odd with
respect to magnetization reversal.
The second term yields
\bege
\begin{aligned}
\label{eq_even_torque_constant_gamma_appendix
}
{\rm t}^{\rm even}_{ij}=&
\frac{e\hbar}{2\pi\mathcal{N}}
\sum_{\vn{k}n\ne m}
{\rm Im}
\left[ 
\langle 
\psi^{\phantom{R}}_{\vn{k}n}  
|
\mathcal{T}_{i}
| 
\psi^{\phantom{R}}_{\vn{k}m}  
\rangle
\langle 
\psi^{\phantom{R}}_{\vn{k}m}  
|
v_{j}
| 
\psi^{\phantom{R}}_{\vn{k}n}  
\rangle
\right]\Biggl\{\\
&\frac{\Gamma
(\mathcal{E}^{\phantom{R}}_{\vn{k}m}
-
\mathcal{E}^{\phantom{R}}_{\vn{k}n})
}{
\left[(\mathcal{E}^{\phantom{R}}_{\rm F}
-
\mathcal{E}^{\phantom{R}}_{\vn{k}n})^2+\Gamma^2\right]
\left[(\mathcal{E}^{\phantom{R}}_{\rm F}
-
\mathcal{E}^{\phantom{R}}_{\vn{k}m})^2+\Gamma^2\right]
}+\\
+&
\frac{
2\Gamma
}
{
\left[
\mathcal{E}^{\phantom{R}}_{\vn{k}n}
-
\mathcal{E}^{\phantom{R}}_{\vn{k}m}
\right]
\left[(\mathcal{E}^{\phantom{R}}_{\rm F}
-
\mathcal{E}^{\phantom{R}}_{\vn{k}m})^2+\Gamma^2\right]
}+\\
+&
\frac{
2
}
{
\left[
\mathcal{E}^{\phantom{R}}_{\vn{k}n}
-
\mathcal{E}^{\phantom{R}}_{\vn{k}m}
\right]^2
}
{\rm Im}\,{\rm ln}
\frac{
\mathcal{E}^{\phantom{R}}_{\vn{k}m}
-
\mathcal{E}^{\phantom{R}}_{\rm F}-i\Gamma
}
{
\mathcal{E}^{\phantom{R}}_{\vn{k}n}
-
\mathcal{E}^{\phantom{R}}_{\rm F}-i\Gamma
}\Biggl\}.
\end{aligned}
\ee
Only ${\rm t}_{ij}^{\rm I(a)}$ 
and ${\rm t}_{ij}^{\rm II}$ contribute 
to ${\rm t}^{\rm even}_{ij}$. From
Eq.~\eqref{eq_time_reversal} it follows 
that ${\rm t}^{\rm even}_{ij}$ is even with respect to
magnetization reversal. 

The odd torkance becomes in the limit $\Gamma\rightarrow 0$
\bege\label{eq_clean_limit_odd_torque_app}
{\rm t}^{\rm odd}_{ij}\overset{\Gamma\rightarrow 0}{=}
\frac{e\hbar}{2\Gamma\mathcal{N}}
\sum_{\vn{k}n}\langle\psi_{\vn{k}n}|\mathcal{T}_{i}|\psi_{\vn{k}n}\rangle
\langle\psi_{\vn{k}n}|v_{j}|\psi_{\vn{k}n}\rangle
\delta(\mathcal{E}^{\phantom{R}}_{\rm F}-\mathcal{E}^{\phantom{R}}_{\vn{k}n}),
\ee
which diverges like $1/\Gamma$.
On the other hand, we obtain in the limit $\Gamma\rightarrow 0$
\bege
{\rm t}^{\rm even}_{ij}\overset{\Gamma\rightarrow 0}{=}
\frac{2e\hbar}{\mathcal{N}}
\sum_{\vn{k}}\sum_{n}^{\rm occ}
\sum_{m\neq n}{\rm Im}
\left[
\frac{
\langle \psi_{\vn{k}n}  |\mathcal{T}_{i}| \psi_{\vn{k}m}  \rangle
\langle \psi_{\vn{k}m}  |v_{j}| \psi_{\vn{k}n}  \rangle
}
{(\mathcal{E}^{\phantom{R}}_{\vn{k}m}-
\mathcal{E}^{\phantom{R}}_{\vn{k}n})^{2}}
\right],
\ee
where the summation over band index $n$ is restricted to the occupied (occ)
bands.
This expression
is independent of $\Gamma$
and thus describes the intrinsic contribution to
the torkance. 

According to Eq.~\eqref{eq_hamiltonian} 
the Hamiltonian $H$ is dependent on the
magnetization direction $\hat{\vn{M}}$ through the exchange interaction
$\mu_{\rm B}\vht{\sigma}\cdot\hat{\vn{M}}\,\Bxc^{\rm xc}(\vn{r})$. The derivative 
of $H$ with respect to magnetization direction $\hat{\vn{M}}$ is
related to the torque operator as follows:
\bege
\hat{\vn{M}}\times \frac{\partial H}{\partial \hat{\vn{M}}}=
\mu_{\rm B}\hat{\vn{M}}\times\vht{\sigma}\,\Bxc^{\rm xc}(\vn{r})=
-\mu_{\rm B}\vht{\sigma}\times\vn{\Bxc}^{\rm xc}(\vn{r})=
\vht{\mathcal{T}}(\vn{r}).
\ee
Using
\bege\label{eq_perturbation_in_k}
\begin{aligned}
\frac{\partial|u_{\vn{k}n}\rangle}{\partial k_j}
&=
\sum_{m\neq n}
\frac{|u_{\vn{k}m}\rangle\langle u_{\vn{k}m}|\frac{\partial H(\vn{k})}{\partial k_{j}}|u_{\vn{k}n}\rangle}
{\mathcal{E}^{\phantom{R}}_{\vn{k}n}-\mathcal{E}^{\phantom{R}}_{\vn{k}m}}\\
&+ia_{\vn{k}nj}|u_{\vn{k}n}\rangle\\
&=
\hbar\sum_{m\neq n}
\frac{|u_{\vn{k}m}\rangle\langle u_{\vn{k}m}|v_{j}(\vn{k})|u_{\vn{k}n}\rangle}
{\mathcal{E}^{\phantom{R}}_{\vn{k}n}-\mathcal{E}^{\phantom{R}}_{\vn{k}m}}\\
&+ia_{\vn{k}nj}|u_{\vn{k}n}\rangle
\end{aligned}
\ee
and
\bege\label{eq_perturbation_theory_for_angles}
\begin{aligned}
\hat{\vn{M}}\times\frac{\partial|u_{\vn{k}n}\rangle}{\partial \hat{\vn{M}}}
=&
\sum_{m\neq n}
\frac{|u_{\vn{k}m}\rangle\langle u_{\vn{k}m}|\hat{\vn{M}}\times\frac{\partial H(\vn{k})}{\partial \hat{\vn{M}} }|u_{\vn{k}n}\rangle}
{\mathcal{E}^{\phantom{R}}_{\vn{k}n}-\mathcal{E}^{\phantom{R}}_{\vn{k}m}}\\
&+i\vn{a}_{\vn{k}n\mathcal{T}}|u_{\vn{k}n}\rangle\\
&=
\sum_{m\neq n}
\frac{|u_{\vn{k}m}\rangle\langle u_{\vn{k}m}|\vht{\mathcal{T}}|u_{\vn{k}n}\rangle}
{\mathcal{E}^{\phantom{R}}_{\vn{k}n}-\mathcal{E}^{\phantom{R}}_{\vn{k}m}}\\
&+i\vn{a}_{\vn{k}n\mathcal{T}}|u_{\vn{k}n}\rangle,
\end{aligned}
\ee
where the phases $a_{\vn{k}nj}$ and $\vn{a}_{\vn{k}n\mathcal{T}}$ determine
the gauge, $H(\vn{k})=e^{-i\vn{k}\cdot\vn{r}}He^{i\vn{k}\cdot\vn{r}}$ 
is the Hamiltonian
in crystal momentum representation 
and $u_{\vn{k}n}(\vn{r})=e^{-i\vn{k}\cdot\vn{r}}\psi_{\vn{k}n}(\vn{r})$ is
the lattice periodic part of the Bloch function $\psi_{\vn{k}n}(\vn{r})$, 
we obtain
\bege
\label{eq_even_torque_clean_limit_app}
{\rm t}^{\rm even}_{ij}\overset{\Gamma\rightarrow 0}{=}
\frac{2e}{\mathcal{N}}
\hat{\vn{e}}_{i} \cdot
\sum_{\vn{k}}
\sum_{n}
\left[
\hat{\vn{M}}\times
{\rm Im}
\left\langle
\frac{\partial u_{\vn{k}n}}{ \partial\hat{\vn{M}} }
\left|
\frac{\partial u_{\vn{k}n}}{\partial k_{j}}\right.
\right\rangle
\right],
\ee
which has the form of a Berry curvature similar to the AHE. 
However, this Berry curvature contribution to the
SOT requires us
to differentiate also with respect to 
the magnetization direction $\hat{\vn{M}}$
and not only with respect to $\vn{k}$. It is thus a mixed
Berry curvature in $\vn{k}$-$\hat{\vn{M}}$ 
space. This mixed $\vn{k}$-$\hat{\vn{M}}$ Berry curvature 
has recently been shown to govern also the 
Dzyaloshinkii-Moriya interaction.~\cite{phase_space_berry,mothedmisot}

\section{Symmetry considerations}
\label{app_symmetry}
General symmetry properties of SOTs are discussed 
in Refs.~\cite{symmetry_spin_orbit_torques,
symmetry_considerations_PhysRevB.88.085423,
symmetry_considerations_PhysRevB.86.094406}.
In the films that we consider in this work
the $xz$ plane is a mirror plane. 
For $\vn{M}$ in $z$ direction, 
mirror reflection at the $xz$ plane inverts $\vn{M}$, because
it is an axial vector. Likewise, the $x$ component of the torque 
is inverted, but the $y$ component
not, because also the torque is an axial vector. 
An electric field along the $x$ direction is not inverted by
mirror reflection at the $xz$ plane, because the electric field
is a polar vector.
Consequently, symmetry requires ${\rm t}_{xx}$ to be 
an odd function of magnetization and
${\rm t}_{yx}$ to be an even function of magnetization 
for $\vn{M}$ in $z$ direction. Thus,
${\rm t}^{\rm even}_{xx}=0$ and ${\rm t}^{\rm odd}_{yx}=0$ if
$\vn{M}$ is in $z$ direction.
When the magnetization is in $z$ direction
the Mn/W(001) films considered in this work exhibit c4
symmetry around the $z$ axis and the Co/Pt(111)-based films
c3 symmetry.
As a consequence of these rotational symmetries we have additionally
${\rm t}^{\phantom{o}}_{yy}={\rm t}^{\phantom{o}}_{xx}={\rm t}^{\rm odd}_{xx}$ 
and 
${\rm t}^{\phantom{e}}_{xy}=-{\rm t}^{\phantom{e}}_{yx}=-{\rm t}^{\rm even}_{yx}$.
Thus, for magnetization along $z$ direction we find the odd torkance
to be a symmetric tensor and the even torkance to be an antisymmetric tensor.
However, in general the even torkance tensor is not always antisymmetric
and the odd torkance tensor is not always symmetric.
Symmetry and antisymmetry of the odd and even parts of the
torkance tensor arise from the 
rotational symmetries of the systems that we consider here 
and not from the
Onsager reciprocity relations of the torkance tensor.

\bibliography{ibcsoit}

\begin{thebibliography}{63}
\bibitem[{com()}]{compare_to_experiment}
\bibinfo{note}{For room temperature measurements on non-annealed
  AlO$_{x}$/Co/Pt trilayers Ref.~\cite{symmetry_spin_orbit_torques} reports
  linear response coefficients for the effective field per current density
  $j_{x}$ of T$_{0}^{\Vert}/(j_{x}\mu_{\rm S})=5\cdot 10^{-7}{\rm mT cm^2/A}$
  for the longitudinal effective field and T$_{0}^{\perp}/(j_{x}\mu_{\rm
  S})=-3.2\cdot 10^{-7}{\rm mT cm^2/A}$ for the transverse effective field. The
  electrical resistivity of the samples is $\rho_{xx}=36\mu\Omega$cm. The
  measured effective fields are the spin-orbit torques per magnetic moment. The
  torkances per magnetic moment are thus t$_{yx}^{\rm
  even}/\mu_{S}$=T$_{0}^{\Vert}$/($\rho_{xx}\,j_{x}\,\mu_{S}$)=0.0139mTcm/V and
  t$_{xx}^{\rm
  odd}/\mu_{S}=$T$_{0}^{\perp}$/($\rho_{xx}\,j_{x}\,\mu_{S}$)=-0.0089mTcm/V.}
\end{thebibliography}

\end{document}